\documentclass[lettersize,journal]{IEEEtran}
\usepackage{amsmath,amsfonts}
\usepackage{algorithmic}
\usepackage{algorithm}
\usepackage{array}
\usepackage[caption=false,font=normalsize]{subfig}
\usepackage{cite}
\usepackage{textcomp}
\usepackage{stfloats}
\usepackage{url}
\usepackage{verbatim}
\usepackage{graphicx}
\usepackage{float}  
\usepackage{subfig}
\usepackage{booktabs}
\usepackage{multirow}
\usepackage{color}
\newtheorem{prop}{Proposition}

\newcommand{\ff}[1] {\textcolor{black}{#1}}
\usepackage[switch,pagewise]{lineno}

\newcommand{\Rmnum}[1]{\uppercase\expandafter{\romannumeral #1}} 
\begin{document}
	
	\title{Retinex Image Enhancement Based on Sequential Decomposition With a Plug-and-Play Framework}
	\author{Tingting Wu, Wenna Wu, Ying Yang, Feng-Lei Fan$^*$,  Tieyong Zeng$^*$
		\thanks{This work was supported in part by the National Key R\&D Program of China under Grants 2021YFE0203700, NSFC/RGC N\_CUHK 415/19, ITF MHP/038/20, CRF 8730063, RGC 14300219, 14302920, and 14301121; in part by CUHK Direct Grant for Research, the Natural Science Foundation of China, under Grants 61971234, 12126340, 12126304, and 11501301; in part by NUPT through the “QingLan” Project for Colleges and Universities of Jiangsu Province, 1311 Talent Plan;  and in part by the Postgraduate Research \& Practice Innovation Program of Jiangsu Province under Grant SJCX21\_0247.}
    \thanks{$^*$Feng-Lei Fan and Tieyong Zeng are co-corresponding authors.}
		\thanks{Tingting Wu and Wenna Wu are with the School of Science, Nanjing University of Posts and Telecommunications, Nanjing (e-mail: wutt@njupt.edu.cn; 1220086613@njupt.edu.cn).}
		\thanks{Ying Yang, Feng-Lei Fan, and Tieyong Zeng are with the Department of Mathematics, The Chinese University of Hong Kong, Shatin, Hong Kong (e-mail: yyang@math.cuhk.edu.hk; hitfanfenglei@gmail.com; zeng@math.cuhk.edu.hk).}}
	
	\markboth{Preprint to IEEE Transactions on Neural Networks and Learning Systems}%
	{Shell \MakeLowercase{\textit{et al.}}: A Sample Article Using IEEEtran.cls for IEEE Journals}
	
	\IEEEpubid{0000--0000/00\$00.00~\copyright~2022 IEEE}
	
	\maketitle
	
	\begin{abstract}
\ff{The Retinex model is one of the most representative and effective methods for low-light image enhancement. However, the Retinex model does not explicitly tackle the noise problem, and shows unsatisfactory enhancing results.
In recent years, due to the excellent performance, deep learning models have been widely used in low-light image enhancement. However, these methods have two limitations: i) The desirable performance can only be achieved by deep learning when a large number of labeled data are available. However, it is not easy to curate massive low/normal-light paired data; ii) Deep learning is notoriously a black-box model \cite{fan2021interpretability}. It is difficult to explain their inner-working mechanism and understand their behaviors. In this paper, using a sequential Retinex decomposition strategy, we design a plug-and-play framework based on the Retinex theory for simultaneously image enhancement and noise removal.
Meanwhile, we develop a convolutional neural network-based (CNN-based) denoiser into our proposed plug-and-play framework to generate a reflectance component.
The final enhanced image is produced by integrating the illumination and reflectance with gamma correction.
The proposed plug-and-play framework can facilitate both post hoc and ad hoc interpretability.
Extensive experiments on different datasets demonstrate that our framework outcompetes the state-of-the-art methods in both image enhancement and denoising. }
	\end{abstract}
	
	
	\section{Introduction}
	\IEEEPARstart{I}{mages} usually exhibit low contrast and unexpected noise distortions when the photographic environment suffers from low illumination. Such drawbacks not only affect the visual quality of captured images but also hinder the effectiveness of downstream vision tasks such as image classification \cite{fan2021sparse, lei2022meta} and segmentation \cite{huang2022learning}. Image enhancement aims to reconstruct a visually pleasing and clear image from its low-light counterpart. During the past decades, various methods have been proposed for low-light image enhancement \cite{pizer1987adaptive, pizer1990contrast, ng2011total, wei2018deep, zhang2019kindling}. Among them, the Retinex model is one of the most representative and significant models. However, most existing optimization-based Retinex models still suffer from image quality degradation such as heavy noise, inadequate details, and low contrast.
	
	The Retinex theory, proposed by Land and McCann \cite{land1971lightness} in 1971, shows an impressive agreement with the color perception of the human visual system (HVS), and inspires numerous image enhancement algorithms \cite{ng2011total, guo2016lime, fu2016weighted}. According to the Retinex theory, an observed image $S$ can be represented by the pixel-wise product of two different components: a reflectance layer $R$ and an illumination layer $L$:
	
	\begin{figure}
		\centering
		\subfloat[Img3]{ \includegraphics[width=0.23\textwidth]{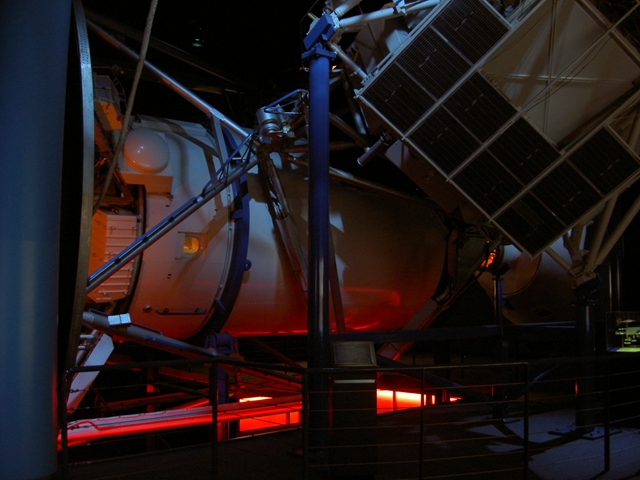}} 
		\subfloat[KinD \cite{zhang2019kindling}]{ \includegraphics[width=0.23\textwidth]{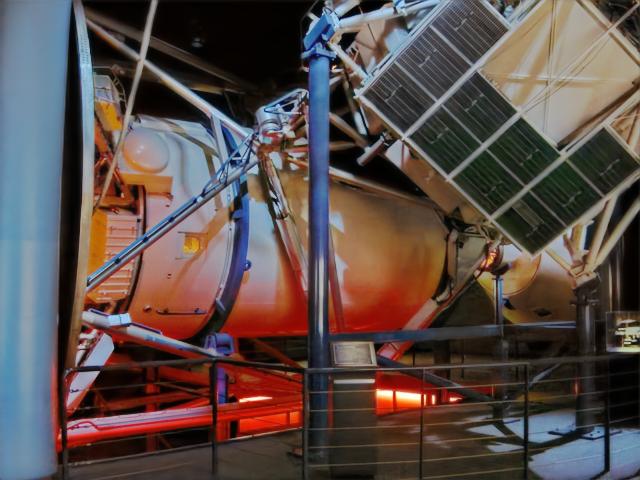}} \\
		\subfloat[FOTV \cite{gu2019novel}]{ \includegraphics[width=0.23\textwidth]{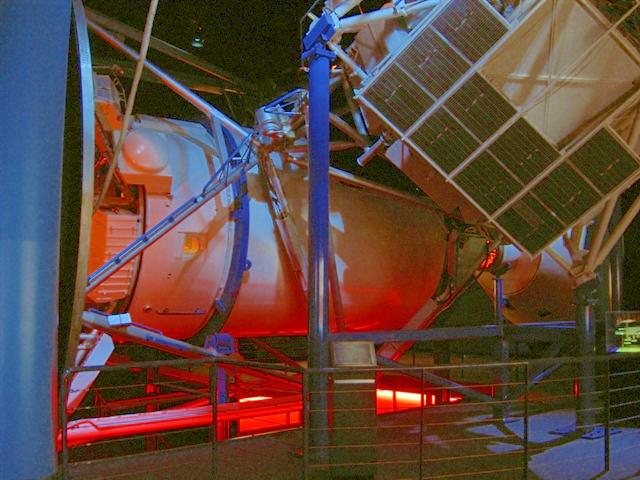}} 
		\subfloat[Ours]{ \includegraphics[width=0.23\textwidth]{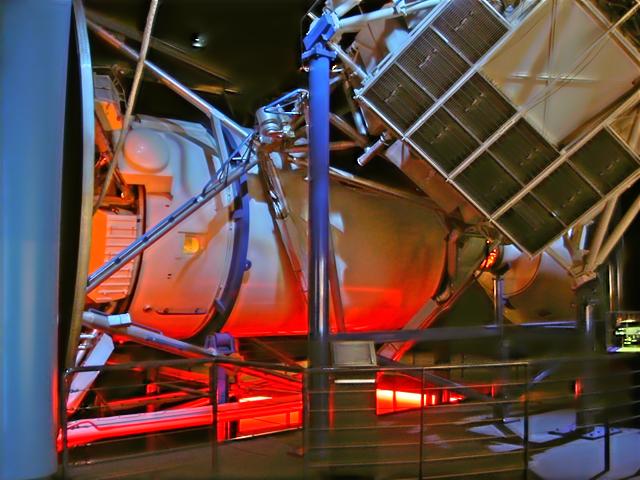}} 
		\caption{An exemplary comparison with two existing methods on Img3. The images generated by KinD \cite{zhang2019kindling} and FOTV \cite{gu2019novel} still have a low contrast. Instead, the image recovered by ours (d) achieves a desirable result.}
	\end{figure}
	
	\begin{equation}
	S=R\cdot L, \label{gs1}
	\end{equation}
	where $R$ denotes the inherent property of the scene surface and contains details and color information of the original image, while $L$ represents the intensity and distribution of the environmental illumination. Note that $L$ is spatially determined by the darkened regions of the image.
	
	\IEEEpubidadjcol

	Simultaneously estimating $R$ and $L$ from $S$ according to Eq. \eqref{gs1} is an ill-posed task. To address this issue, priors are needed to be incorporated. 
	Michael \emph{et al}. \cite{ng2011total} adopted a log-transform to overcome ill-posedness of the Retinex model. Then, the total variation regularization and $L_2$-norm regularization were employed to estimate the log-reflectance component and log-illumination component, respectively. 
	However, the log-transform may seriously distort gradient information, which is problematic for noise in image enhancement tasks \cite{ng2011total, chang2015retinex, fu2016weighted}.
	To circumvent the above-mentioned limitations, numerous decomposition algorithms have been investigated without the log-transform.
	Gu \emph{et al}. \cite{gu2019detail} predicted the illumination and the reflectance directly in the image domain by $L_2$ and $L_1$ norms, respectively.
	Following the work in \cite{gu2019detail}, Gu \emph{et al}. \cite{gu2019novel} proposed a Retinex-based fraction-order total variational (FOTV) model by employing the fraction-order gradient total variation regularization  $\left( \nabla^{\alpha}, \alpha \in \left( 1,2 \right)\right)$ on both the reflectance and the illumination. However, such a decomposition method is mainly carried out in the V-channel of the HSV color space, which ignores the noise in other two channels. 
	Thus, although these methods can remove noise to a certain extent while preserving finer details, their denoising ability is still limited.

	\begin{figure}
		\centering
		\subfloat[]{ \includegraphics[height=0.1\textwidth]{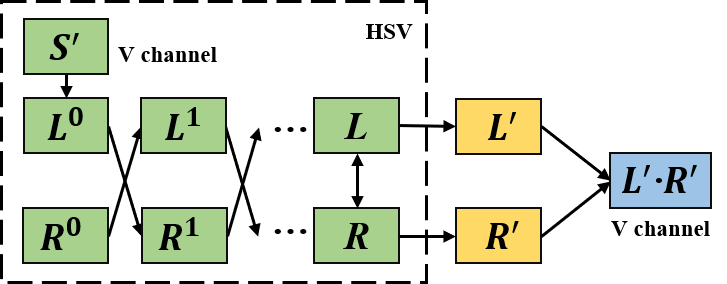}} 
		\subfloat[]{ \includegraphics[height=0.1\textwidth]{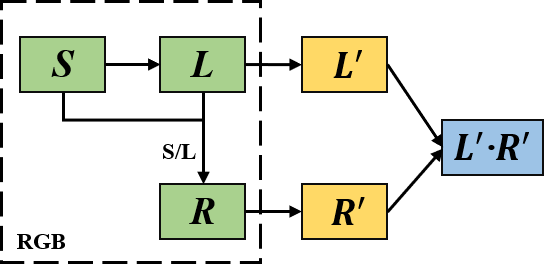}} 
		\caption{Comparison between the common and our sequential decomposition strategies.  The parts in the dotted boxes represent the Retinex decomposition operations. (a) shows the alternating decomposition strategy, and (b) is our Retinex sequential decomposition method. $L'$ and $R'$ demonstrate the illumination and reflectance obtained by gamma correction.}
		\label{flo} 
	\end{figure}

	\ff{It is increasingly noticed that noise in the dark channel is a key issue in image enhancement, which must be treated appropriately \cite{wei2018deep,zhang2019kindling, lv2021attention, xu2020learning,zhu2020eemefn, chen2018learning, wang2019progressive}. To desirably enhance the low-light illuminated images, the model should be endowed with denoising capability. In recent years, due to its excellent performance, deep learning has been widely used in low-light image enhancement. Realizing the criticality of denoising, a plethora of deep models \cite{wei2018deep,zhang2019kindling, lv2021attention, xu2020learning,zhu2020eemefn, chen2018learning, wang2019progressive} take the denoising as a separate module. However, these methods have two limitations: i) The desirable performance can only be achieved by deep learning when a large number of labeled data are available. However, it is not easy to curate massive low/normal-light paired data due to the following reasons. First, there is no convincing synthetic method to simulate realistic low-light images from normal-light ones because illumination conditions are highly varied. Second, the existing datasets are either small or collected in a too specific context such as the extremely underexposed condition to fulfill the general need. Then, the model trained over such datasets may not be translatable; ii) Deep learning is notoriously a black-box model \cite{fan2021interpretability}. Despite that deep learning delivers great performance in real-world tasks, it is difficult to explain their inner-working mechanism and understand their behaviors. } 
 
 \ff{To address these issues, inspired by the plug-and-play image restoration method \cite{venkatakrishnan2013plug}, we propose a framework that uses sequential decomposition strategy to solve $R$ and $L$ sequentially, thereby avoiding the alternating iteration and canceling the mutual interference between solving $R$ and $L$. Fig. \ref{flo} exhibits differences between the common Retinex decomposition strategy and the sequential decomposition strategy. This framework circumvents the reliance on large paired low/high-light data, a key problem encountered in image enhancement, by transforming the problem of learning how to enhance an image into the problem of learning how to do image denoising in solving $R$. The training of a CNN-based denoiser can be done reliably and effectively over synthetic noisy/clean data because synthesizing noise is much easier and more faithful than synthesizing light illumination. Furthermore, the plug-and-play framework promotes interpretability. On the one hand, due to the modularized structure of the plug-and-play framework, it is relatively easier to apply post hoc analysis to identify the mechanism of the framework. Thus, we can make the most use of the performance of advanced deep learning denoisers and use post hoc analysis method to gain interpretability. Along this direction, instead of translating the existing CNN denoisers into this task, we independently design a simple yet efficient and effective CNN-based denoiser. On the other hand, we can directly seek an explainable CNN-based denoiser. The employment of an interpretable denoiser can further enhance the entire framework's interpretability. Specifically, we use a wavelet-inspired autoencoder \cite{fan2020soft}, which is essentially a learnable wavelet shrinkage model.
 Overall, our contributions are threefold:}
	
	\begin{enumerate}
		\item \ff{We present an efficient sequential decomposition Retinex algorithm to solve the illumination and reflectance function, respectively. To the best of our knowledge, it is the first time to investigate the problem of low-light image enhancement by a novel plug-and-play framework. This framework is an organic fusion of model-driven and data-driven modalities, which can circumvent the reliance on large low/normal-light data. }
		
		\item \ff{Through a post hoc analysis and applying an explainable denoiser, we show how the modularized structure in the proposed plug-and-play framework facilitates interpretability in terms of both post hoc analysis and ad hoc interpretable modeling.}
		
		
		\item Extensive and systematic low-light image enhancement experiments are conducted on different datasets to demonstrate the performance of our proposed method. The comparisons have demonstrated that the proposed model outperforms the existing methods from both quantitative and qualitative aspects.
	\end{enumerate}

	\section{Related Work} \label{related_work}
	\subsection{Low-Light Image Enhancement Methods}
	During the past decades, a large number of algorithms have been developed for low-light image enhancement. These algorithms can be reasonably categorised as model-based and learning-based algorithms.
	
	\subsubsection{Model-Based Methods}
	In this part, we mainly introduce two model-based methods: the histogram equalization (HE)-based methods and the decomposition-based methods.
	
	Intuitively yet effectively, HE enhances image contrast by changing the histogram distribution. The basic principle of HE and its variants is to expand the dynamic range of pixels in an image. Different from the standard histogram equalization algorithm, adaptive histogram equalization (AHE) \cite{pizer1987adaptive} adjusts image contrast by calculating the local histogram and redistributing the brightness. Although AHE enhances local details in extremely dark or bright areas, it also amplifies noise. The work in \cite{pizer1990contrast} can alleviate overexposure and noise amplification by limiting the distribution of gray levels in different regions (CLAHE).
	Subsequently, various HE-based methods are devised to improve the overall visual quality based on different constraints \cite{arici2009histogram,ibrahim2007brightness}.
	
	Unfortunately, these HE-based methods are handicapped to images with nonuniform illumination, which motivates the invention of decomposition-based methods. This thread of work assumes that an image can be decomposed into the reflectance and the illumination components, and an image is enhanced by further processing and integrating these two components.
	The existing decomposition-based approaches can be categorised into the variational methods \cite{ng2011total, chang2015retinex, ren2020lr3m, guo2016lime, gu2019novel, fu2016weighted, gu2019detail, wang2021image, liu2018retinex, liang2015retinex}, path-based algorithms \cite{cooper2004analysis, provenzi2005mathematical}, recursive algorithms\cite{land1986alternative, mccann1999lessons}, partial differential equations (PDE) based methods \cite{horn1974determining, morel2010pde}, and learning-based methods \cite{wei2018deep, zhang2019kindling, liu2021underexposed, chen2018learning, ma2021learning}. However, these techniques tend to induce sketchy details, prominent noises, and unknown artifacts.

	\subsubsection{Learning-Based Methods}
	Recently, many learning-based methods have emerged with promising performance for underexposed image correction. The key bottleneck therein is the lack of low/high-light datasets. To solve this problem,
	Wang \emph{et al.} \cite{wei2018deep} collected a paired dataset named LOL, which was widely applied in many works. Meanwhile, they introduced an end-to-end trainable network called RetinexNet, which is the first work that combines the Retinex theory and deep learning.
	\ff{Chen \emph{et al.}  \cite{chen2018learning} built the SID dataset and trained an end-to-end network with good noise reduction and image enhancement, particularly for the extremely underexposed images.}
	Considering the poor generalization performance of these methods in other test sets, some work attempts to adopt semi-supervised or unsupervised learning, such as RetinexDIP \cite{zhao2021retinexdip}. However, their results sometimes suffer from unexpected artifacts and improper exposure.
	
	Another key problem in the image enhancement task is the noise hidden in the dark. Lore \emph{et al.} \cite{lore2017llnet} exploited an autoencoder to extract the image information and achieved the purpose of enhancement. Wang \emph{et al.} designed the DeepUPE \cite{wang2019underexposed}, which establishes the loss function according to the prior information of illumination. Since these methods are not designed for noise, the enhanced results still suffer visible noise.
	Some learning-based methods treat denoising as an independent module. \ff{Lv \emph{et al.} \cite{lv2021attention} proposed a multi-branch decomposition-and-fusion enhancement network to cope well with color distortion and noise.
	Xu \emph{et al.} \cite{xu2020learning} designed a frequency-based network for denoising and enhancement simultaneously.
	They adopted a CDT module to eliminate noise and protect details.
	Zhu \emph{et al.} \cite{zhu2020eemefn} designed a two-stage network to effectively remove noise through the multi-exposure fusion module. In our study, we also regard noise suppression as a non-negligible factor. Progressive Retinex \cite{wang2019progressive} uses two fully pointwise convolutional neural networks to respectively simulate the statistical regularities of ambient light and image noise, and leverage them as constraints to facilitate the mutual learning process, which not only avoids the ambiguity between tiny textures and image noise but also enhances computational efficiency. Liu \textit{et al.} \cite{liu2021retinex} focused on constructing a lightweight yet effective network, referred to as RUAS \cite{liu2021retinex} to enhance low-light images in real-world scenario. Yang \textit{et al.} \cite{yang2021sparse} noticed that there lacks a desirable objective for low-light image enhancement, and they designed an end-to-end signal prior-guided layer separation network with layer-specified constraints. }

    \ff{Despite promising performance, these methods either are trained over synthetic data/data collected in a specific condition or suffer from the lack of interpretability. In contrast, by synergizing the model-driven and data-driven modalities, our plug-and-play framework delivers superior performance without dependence on paired low/normal-light data and enjoys interpretability.}

	\subsection{The Plug-and-Play Framework}
	Earlier, the plug-and-play framework was proposed to solve the denoising problem \cite{venkatakrishnan2013plug}, which allows the insertion of different denoisers for prior knowledge learning. The plug-and-play framework mainly contains two steps. First, the objective function is decoupled into a fidelity subproblem and a prior subproblem via the variable splitting algorithms. These two subproblems are combined into an iterative scheme solved alternately. Second, the existing state-of-the-art denoising techniques can be employed directly to solve the prior subproblem, including regularization denoising methods \cite{gu2014weighted, huang2021quaternion} and learning-based denoising methods \cite{fang2020multilevel}. 
	
	Recently, various image reconstruction approaches have been explored based on the plug-and-play framework, such as image super-resolution \cite{zhang2019deep}, image deblurring \cite{zhang2017learning} and image denoising \cite{zhang2017learning, sun2019block}. Zhang \emph{et al.} introduced the IRCNN \cite{zhang2017learning} model for non-blind image deblurring and image denoising by plugging a deep denoiser prior into the half quadratic splitting (HQS) algorithm \cite{geman1995nonlinear}.
	Sun \emph{et al.} \cite{sun2019block} developed a block-coordinate regularized denoising algorithm, which decomposes the large-scale estimation problem into a series of updates covering a small part of unknown variables. These methods have revealed the surprising potential of the plug-and-play framework in different image restoration tasks.
	
	One of the major advantages of the plug-and-play framework is that a pre-trained denoiser can be used if there is no enough data for end-to-end training. Since the available paired data for image enhancement tasks are very limited, the plug-and-play framework is suitable for our task. At the same time, the model can gain interpretability from a mathematical perspective, since there is a close tie between the plug-and-play framework and the traditional restoration methods.

	\section{Model and Algorithm} \label{algorithm}
	
	
	Our proposed method mainly consists of three components: the sequential decomposition module, the denoising module, and the adjustment module. Because of the sequential decomposition, $L$ and $R$ are solved independently. For the illumination layer $L$, we first compute the initial illumination $\hat{L}$ from the input $S$ via the meanRGB operation; then, we derive $L$ by minimizing a loss function through iterative optimization. While for the reflectance layer $R$, since $L$ is done, we can get the initial map of $R$ through the simple relationship, \textit{i.e.}, $R^{(0)}=S/L$. Then, the reflectance map $R$ is processed by a CNN-based denoiser. Finally, the Gamma correction is employed to combine the illumination and reflectance maps to yield the final enhanced image. Fig. \ref{flow} illustrates the flowchart of our proposed method based on the sequential decomposition within a plug-and-play framework. 
	
	\subsection{Retinex Model With the Plug-and-Play Scheme}
	We propose a plug-and-play scheme for low-light image enhancement with the sequential decomposition-based Retinex model. We construct the following optimization function to alleviate the ill-posedness of Eq. (\ref{gs1}):
	
	\begin{equation}
	\underset {R, L} {\text{min}}~\Vert S-L \cdot R\Vert_F^2+\lambda\Phi\left(R\right)+\beta\Vert\nabla R-G\Vert^2_F+\alpha\Vert \nabla L \Vert_1,
	\label{R}
	\end{equation}
	where $S$, $L$, and $R$ denote the low-light image, estimated illumination and reflectance, respectively, $G$ is the adjustment to $\nabla S$, $\lambda$, $\alpha$, $\beta$ are the regularization parameters, $\Vert\cdot\Vert_F$ represents the Frobenius norm, $\Vert\cdot\Vert_1$ denotes the $\ell_1$ norm. Next, we explain in detail each term in the objective function: 
	
	\begin{figure*}[t]
		\centering
		\includegraphics[scale=0.55]{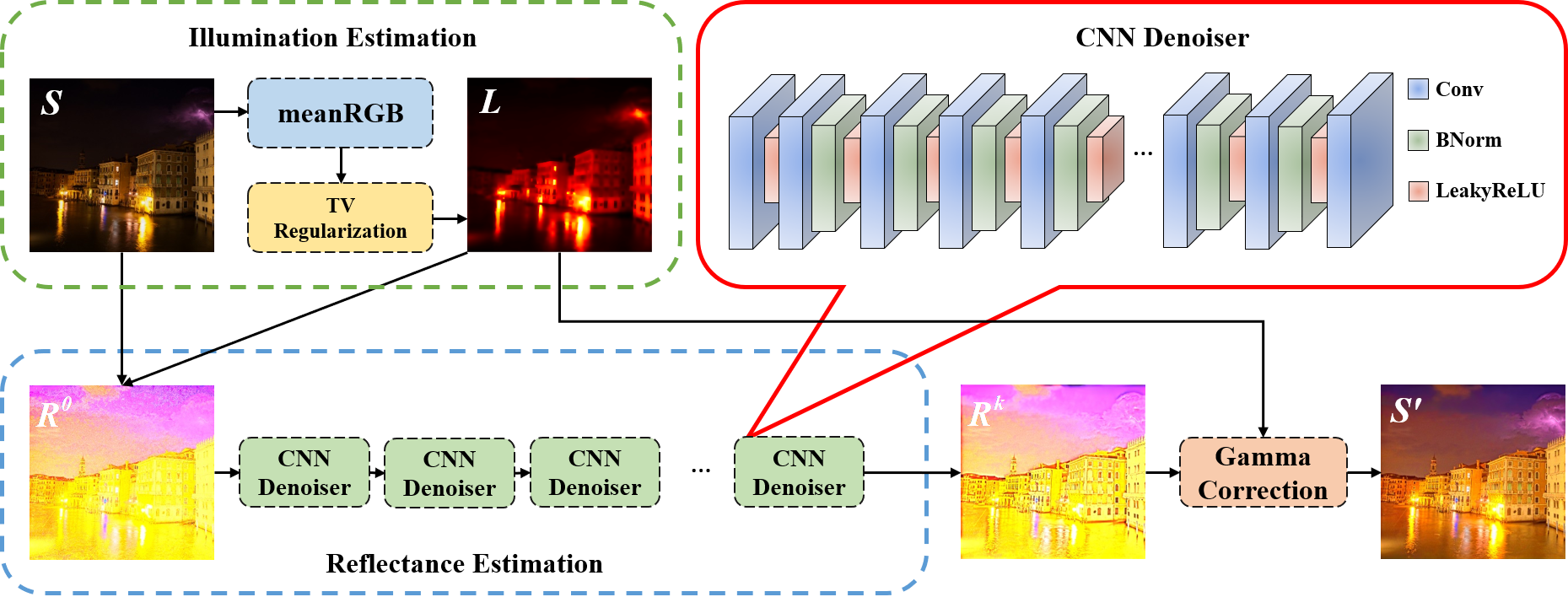}
	\caption{The flowchart of our method. We first obtain the illumination $L$ via refining the initial illumination $\hat{L}$ generated by the meanRGB operation. Then we obtain the initial value of $R$ via $R^{(0)}=S/L$. The noise-free reflectance map is obtained by inserting the denoiser with the designed plug-and-play framework. Finally, gamma correction is applied to the decomposed component to adjust the image.}
		\label{flow}
	\end{figure*}
	
	\begin{itemize}
		
		\item  The term $\Vert S-L \cdot R\Vert_F^2$ minimizes the Frobenius distance between $S$ and $L\cdot R$, which is the fidelity term. 
		\item  $\Phi\left(R\right)$ is the regularization term of $R$, which is used to denoise the reflectance part. If $\Phi\left(R\right)=\Vert\nabla R\Vert_1$, the proposed model (\ref{R}) is a hybrid model about total variation. If $\Phi\left(R\right)=\sum_i\Vert NN_i(R)\Vert_*$, the proposed model (\ref{R}) is a low-rank hybrid model, here $\Vert\cdot\Vert_*$ represents the  nuclear norm, and $NN_i(R)=[R_{i1},R_{i2},...,R_{ik}]$ is the similar patch group of the reference patch $R_{i1}$.
		In our proposed framework, $\Phi\left(R\right)$ is an implicit prior whose subproblem is denoising.
		\item The third term $\Vert\nabla R-G\Vert_F^2$ minimizes the distance between $\nabla R$ and $G$. Since 
		$G$ is obtained from $\nabla \hat{S}$ (adjusted by the gradient of $S$), the details of edges can be preserved. $\nabla \hat{S}$ and $G$ \cite{li2018structure} is given by
		\begin{equation}
		\nabla \hat{S} = \begin{cases}
		0,&{\mathrm {if}}\ {\nabla\hat{S}<\varepsilon}\\
		{\nabla S,}&{\mathrm {otherwise}}
		\end{cases}
		\label{9}
		\end{equation}
		\begin{equation}
		G=(1+\kappa e^{-\vert\nabla \hat{S}\vert/\sigma})\circ \nabla \hat{S}.
		\label{G}
		\end{equation}
		Here, $\epsilon$ can filter out small gradients; $\sigma$ and $\kappa$ are the parameters that control the enlarged level. Eq. (\ref{9}) can suppress the possible noise before the amplification.
		
		\item $\Vert\nabla L\Vert_1$ ensures the smoothness of the illumination layer.
	\end{itemize}
	

	
    Typically, to estimate the illumination map $L$ and the reflectance map $R$, we can solve Eq. (\ref{R}) in an alternating fashion (solving $L$ and $R$ alternatively in the iteration). But this is time-consuming and complex. To address this issue, we adopt a novel sequential decomposition strategy to generate the illumination part independently. Specifically, we first estimate the illumination $L$ from the initial illumination $\hat{L}$ (Eq. (\ref{L})), and then estimate the reflectance map $R$ (Eq. (\ref{subR})):
    
	
	\begin{equation}
	\underset {L} {\text{min}}~\Vert L-\hat{L}\Vert_F^2+\alpha\Vert \nabla L \Vert_1
	\label{L}
	\end{equation}
	\begin{equation}
	\underset {R} {\text{
			min}}~\Vert S-L \cdot R\Vert^2_F+\lambda\Phi\left(R\right)+\beta\Vert\nabla R-G\Vert^2_F,
	\label{subR}
	\end{equation}
	
	\subsubsection{Illumination Map ($L$) Estimation}
	Recently, numerous approaches have been proposed to directly estimate the initial illumination map $\hat{L}$, such as the maxRGB \cite{land1977retinex} or meanRGB \cite{ren2020lr3m} operator.
	These approaches pre-assume that each channel of an RGB image has a common illumination. We adopt the meanRGB operator to ensure the consistency of illumination, which is defined as
	\begin{equation}
	\hat{L}(x)=\frac{1}{3}\sum_{c\in\{R,G,B\}}S^c(x),
	\label{LH}
	\end{equation}
	where $x$ denotes the pixel of the image. 
	Then, an alternating direction minimization method (ADMM) \cite{boyd2011distributed} is employed to solve Eq. $\left(\ref{L} \right)$:
	\begin{equation}
	\begin{aligned}
	&\underset {L} {\text{
			min}}~\Vert L-\hat{L}\Vert_F^2+\alpha\Vert v \Vert_1, \quad\text{ s.t. }\nabla L=v,
	\end{aligned}
	\end{equation}
	where $v$ is an auxiliary variable. Then, we can create a corresponding augmented Lagrangian function:
	\begin{equation}
	\begin{aligned}
	\mathcal{L}_\theta(L, v)=\Vert L-\hat{L}\Vert_F^2+\alpha\Vert v \Vert_1&\\+\frac{\theta}{2}\Vert \nabla L-v\Vert_F^2&+\langle Z, \nabla L-v\rangle,
	\end{aligned}
	\end{equation}
	where $\theta$ is the penalty parameter, and $Z$ is the Lagrangian multiplier. The ADMM algorithm is derived by minimizing $\mathcal{L}$ with respect to $L$ and $v$ (one at a time while fixing the other at its most recent value).
	By direct computation, we can get the updating formula of $L$:
	\begin{equation}
	\begin{aligned}
	L^{(k+1)}=(2+\theta^{(k)}\nabla^{T}\nabla)^{-1}&\\(2\hat{L}+&\theta^{(k)}\nabla^{T}v^{(k)}-\nabla^{T}Z^{(k)}),
	\end{aligned}
	\label{11}
	\end{equation}
	which is implemented by fast Fourier transform (FFT) and inverse fast Fourier transform (IFFT).	
	Meanwhile, the $v$-subproblem can be quickly solved by the soft shrinkage:
	\begin{equation}
	v^{(k+1)}=\eta(\nabla L^{(k+1)}+\frac{Z^{(k)}}{\theta^{(k)}},\frac{\alpha}{\theta^{(k)}}),\label{soft}
	\end{equation}
	where $\eta$ is a soft shrinkage function defined as
	\[
	\eta(x_{ij},c):=\frac{x_{ij}}{\vert x_{i,j} \vert}\cdot \max (\vert x_{i,j} \vert-c,0).
	\]
	Finally, the Lagrangian multiplier $Z$ and the parameter $\theta$ are updated through
	\begin{equation}
	\begin{aligned}
	&Z^{(k+1)}\gets Z^{(k)}+\theta^{(k)}(\nabla L^{(k+1)}-v^{(k+1)})\\
	&\theta^{(k+1)}\gets\theta^{(k)}\rho,\quad \rho >1.
	\label{nu}
	\end{aligned}
	\end{equation}
	Here $\rho$ is the step size. The iteration will terminate when $\Vert\nabla L-v\Vert_F\leq\iota\Vert\hat{L}\Vert_F$ with $\iota=10^{-5}$ or $k$ reaches the maximum value. 
	
	\subsubsection{Solution of R-subproblem}
	Following the half quadratic splitting method \cite{geman1995nonlinear}, Eq. (\ref{subR}) can be reformulated as a constrained optimization problem by introducing an auxiliary variable $z$:
	\begin{equation}
	\underset {R} {
		\min}~\Vert S-L \cdot R\Vert^2_F+\lambda\Phi\left(z\right)+\beta\Vert\nabla R-G\Vert^2_F,\\
	\quad\text{ s.t. }R=z.
	\label{subR2}
	\end{equation}	
	Then Eq. (\ref{subR2}) is solved by minimizing the following problem:
	\begin{equation}
	\begin{aligned}
	\mathcal{L}_\mu(R,z)=\Vert S-L \cdot R\Vert^2_F+\lambda\Phi\left(R\right)&\\+\beta\Vert\nabla z-G\Vert^2_F&+\frac{\mu}{2}\Vert z-R \Vert^2_F,
	\label{La}
	\end{aligned}
	\end{equation}
	where $\mu$ is a positive penalty scalar. 
	
	\underline{$z$-subproblem (contrast enhancement):}
	The fidelity term and regularization term are decoupled into two individual subproblems. Collecting the $z$-involved terms from Eq. ($\ref{La}$) gives the problem as follows:
	\begin{equation} 
	z^{(k+1)} = \underset {z} {
		\arg\min}~ \beta\Vert\nabla z-G\Vert^2_F+\frac{\mu}{2}\Vert R-z^{(k)} \Vert^2_F.\label{Ra}
	\end{equation}	
	Then the updating formula of $z^{k+1}$ is given by
	
	\begin{algorithm} [t]
		\caption{Single-Image Low-Light Enhancement}  
  \label{al}
		\begin{algorithmic}[1]
			\REQUIRE The input image $S$, $k=0$, \ff{$\alpha=0.1$, $\theta=0.0045$, $\rho=1.08$, $Z^{(0)}=0$, $\epsilon=1$, $\kappa=2.5$, $\sigma=10$;}
			\STATE Compute $G$ via Eqs. $\left(\ref{9}\right)$ and $\left(\ref{G}\right)$;
			\STATE Estimate $\hat{L}$ via Eq. $\left(\ref{LH}\right)$;
			\WHILE{not converged}
			\STATE Update $L^{\left(k+1\right)}$ via Eq. $\left(\ref{11}\right)$;
			\STATE Update $v^{\left(k+1\right)}$ via Eq. $\left(\ref{soft}\right)$;
			\STATE Update $Z^{\left(k+1\right)}$ via Eq. $\left(\ref{nu}\right)$;
			\ENDWHILE  
			\REQUIRE $R^{\left(0\right)}=S/L$, $k=0$,  \ff{$\mu=0.001$, $\beta=0.001$, noise level $\sqrt{\lambda}$, gamma correction coefficient $\gamma_1$,$\gamma_2$;}
			\WHILE{not converged}
			
			\STATE Update $z^{\left(k+1\right)}$ via Eq. $\left(\ref{uu}\right)$;
			\STATE Update $R^{\left(k+1\right)}$ via Eq. $\left(\ref{zz}\right)$;
			\ENDWHILE  
			\STATE  Estimate $S^{\prime}$ via Eq. $\left(\ref{gamma}\right)$.
			\ENSURE The estimated result $S^{\prime}$.      
		\end{algorithmic}  
	\end{algorithm}	
	\begin{equation}
	z^{(k+1)}=\frac{2\beta\nabla^{T}G+\mu R^{(k)}}{2\beta\nabla^{T}\nabla+\mu I}.
	\label{uu}
	\end{equation}
	Note that the update of $z^{(k+1)}$ can be implemented by FFT and I{FF}T.
	
	\underline{$R$-subproblem (noise suppression):} Neglecting the terms unrelated to $R$, the $R$-subproblem can be solved by the following iterative scheme:
	
	\begin{equation}
 	\begin{aligned}
	    &\underset {R} {
		\arg\min}~\Vert S-L \cdot R\Vert^2_F+\frac{\mu}{2}\Vert z^{(k+1)}-R \Vert^2_F+\lambda\Phi\left(R\right)\\
          \Rightarrow&\underset {R} {
		\arg\min}~\frac{L^2}{\lambda}\Vert R-\frac{S}{L}\Vert_F^2+\frac{\mu}{2\lambda}\Vert R-z^{(k+1)}\Vert_F^2+\Phi\left(R\right)\\
        \Rightarrow&\underset {R} {
		\arg\min}~\frac{1}{2(\sqrt{\lambda})^2}\left\| R-\frac{2S\cdot L+\mu z^{(k+1)} }{2L^2+\mu  I}\right\|^2_F+\Phi\left(R\right),
  \label{Rb}
  	\end{aligned}
	\end{equation}
 Based on Bayesian statistics, we use a Gaussian denoiser to help reconstruct the component $R
 $ and rewrite Eq. $\left(\ref{Rb}\right)$ as
	\begin{equation}
R^{(k+1)}=\mathrm{Denoiser}(\frac{2S\cdot L+\mu z^{(k+1)} }{2L^2+\mu I},\sqrt{\lambda}).
	\label{zz}
	\end{equation}

\ff{\textbf{Summary}: According to the Retinex theory, an observed image $S$ can be represented by the pixel-wise product of two different components: a reflectance layer $R$ and an illumination layer $L$:
	$S=R\cdot L$,
where $R$ denotes the inherent property of the scene surface and contains details and color information of the original image, while $L$ represents the intensity and distribution of the environmental illumination. 
Our framework uses a sequential decomposition strategy to solve $R$ and $L$ sequentially, thereby avoiding the alternating iteration and canceling the mutual interference between solving $R$ and solving $L$. Fig. \ref{flo} exhibits differences between the common Retinex decomposition strategy and the sequential decomposition strategy. Furthermore, the important prior subproblem in solving $R$ is reduced to a denoising problem, amendable to the advanced denoisers to be employed in a plug-and-play manner. This is why our framework is plug-and-play.}
 
\subsection{The CNN-based Denoisers}

	\ff{Since our framework is plug-and-play, denoisers can be selected as appropriate. In recent years, deep learning has been widely used in various fields \cite{ma2021dirichlet,du2021progressive}. Considering that learning-based methods significantly outperform traditional variational models, we design a CNN-based denoiser to solve the $R$ subproblem in image enhancement task. In addition, we also use an explainable denoiser.}
  
\ff{\underline{Ours}: Currently, the activation unit ReLU is the most widely-used nonlinear activation function in deep learning because of its good properties in preventing gradients vanishment or explosion. However, ReLU also tends to block information transmission because it sets zero to all negative parts of the input, which may block reasonable information circulation because those inhibited neurons will not be updated in the backpropogation. 
To address this issue, a thread of the activation functions, \textit{e.g.}, ON/OFF ReLU \cite{kim2015convolutional}, Concatenated-ReLU \cite{shang2016understanding}, and Leaky-ReLU \cite{maas2013rectifier}, that consider the information flow of negative parts were proposed to supplement ReLU in deep networks. Compared to other activation functions, Leaky-ReLU has a smaller number of parameters and a reasonably effective generalization performance. Therefore, we design the CNN-based denoiser with Leaky-ReLU, as shown in Fig. \ref{flow}.  
	The new denoiser mainly includes three parts: ``Conv+LeakyReLU" for the first layer, ``Conv+BactchNorm+LeakyReLU" for the 2$\sim$18 layers, ``Conv" for the last layer. 
	The convolutional layer in the first part uses 64 filters to generate 64 feature maps and the size is $3\times3\times 3$. And we use the filters of size $3\times3\times 64$ for the second and last parts. The LeakyReLU function is defined as follows:
	\[
	\mathrm{LeakyReLU}(x) =
	\begin{cases}
	x,& {x>0}\\
	{a x,}&{x\geq0}\\
	\end{cases},
	\]
where $a$ denotes a small constant, and is usually set to $-0.02$. We also add batch normalization between the convolutional layer and the activation function. The application of batch normalization can also improve training efficiency and denoising performance \cite{ioffe2015batch}.}

\begin{figure}
		\centering
		\subfloat[]{ \includegraphics[height=0.25\textwidth]{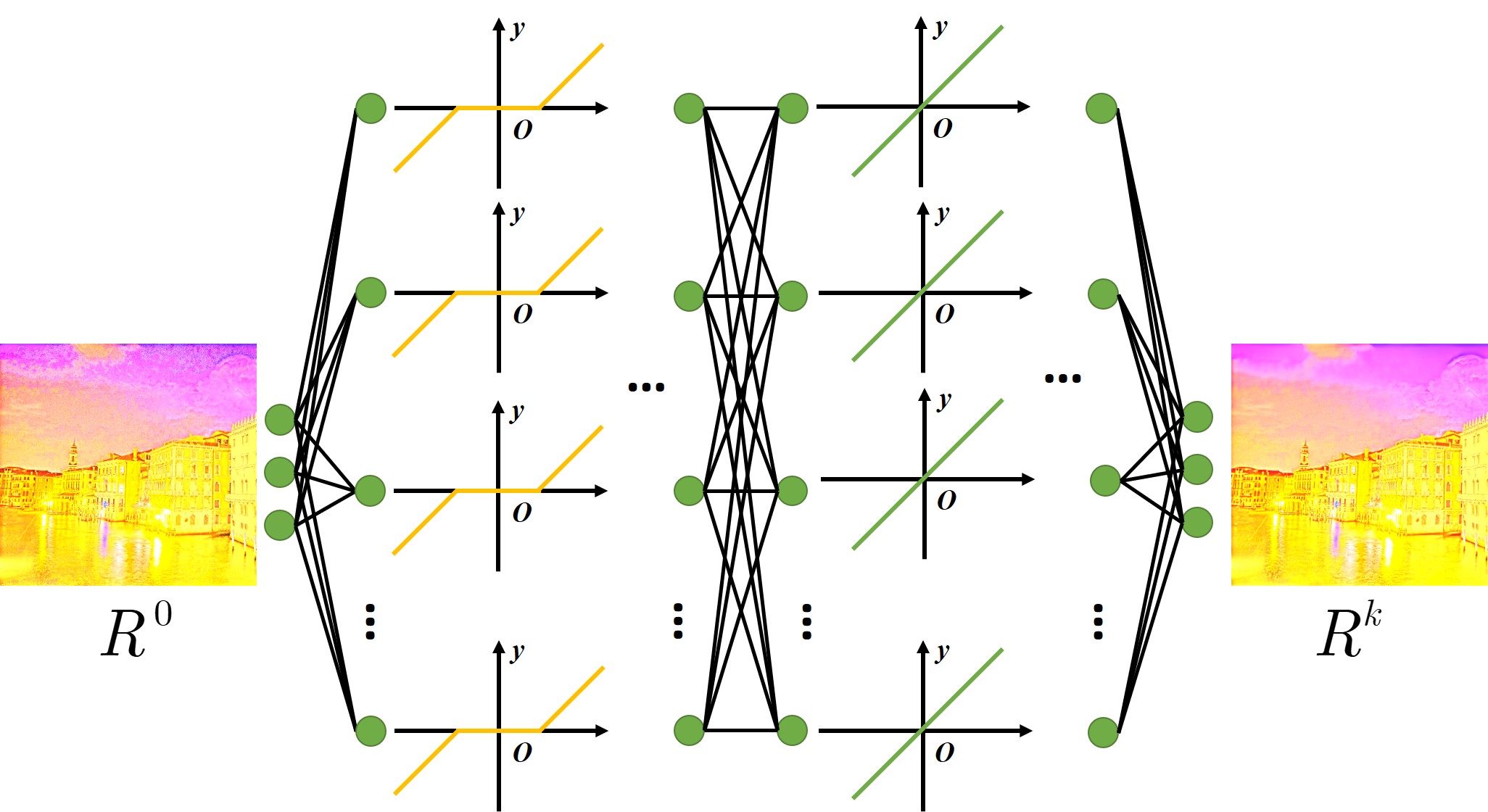}} 
		\caption{The architecture of Soft-AE. The Soft-AE is a learnable wavelet shrinkage algorithm. } 
		\label{afunc}
	\end{figure}

 \underline{Soft-AE (An Explainable Denoiser)}: \ff{Fan et al. \cite{fan2020soft} proposed the so-called soft autoencoder (Soft-AE), which is interpretable based on the wavelet shrinkage theory. As shown in Fig. \ref{afunc}, the activation functions in the encoding layers are set to soft-thresholding units $\eta_{b<0}(x)=\mathrm{sgn}(x)\max\{|x|+b,0\} $, where $b$ is a threshold, $\mathrm{sgn}(\cdot)$ is the sign function, and activation functions of the decoding layers are linear. As a result, convolutional layers in the encoder part conduct wavelet transform, while convolutional layers in the decoder part conduct inverse wavelet transform. The Soft-AE can be regarded as unrolling the cascade wavelet shrinkage algorithm into a network, and the wavelet transform is learnable.}

 \ff{Mathematically, the wavelet shrinkage algorithm consists of three steps. Suppose that we have the following additive noise model:  $Y(t)=S(t)+N(t)$, where  $Y(t)$  and  $S(t)$  are the measured and the authentic signals, respectively. (a) perform the wavelet transform over the noise signal to derive wavelet coefficients: $\hat{Y}=W[Y] $; (b) apply an element-wise soft-thresholding activation to the wavelet coefficients: $Z=\eta_{-\sigma_{N} \sqrt{2 \log n}}(\hat{Y}) $, where    is the noise variance, and  $n$  is the number of pixels; (c) perform the inverse wavelet transform:  $\hat{S}=W^{-1}[Z]$.}

\begin{algorithm} [h]
\caption{Wavelet Shrinkage Algorithm}  
\begin{algorithmic}[1]
\REQUIRE $Y(t)=S(t)+N(t)$, wavelet $\phi$
\STATE Wavelet transform by $\phi$: $\hat{Y}=W_{\phi} [Y(t)]$;
\STATE Soft thresholding: $Z=\eta_{b}(\hat{Y})$;
\STATE Inverse wavelet transform by $\phi^{-1}$: $\hat{S}=W_{\phi}^{-1}[Z]$;
\ENSURE $\hat{S}(t)$
\end{algorithmic}  
\label{alg:wsa}
\end{algorithm}	

\begin{prop}
    Let $\hat{S}(t)$ and $S(t)$ be the recovered signal and the noisy signal in Algorithm \ref{alg:wsa}. Given the Besov norm $B$ that is a measure for smoothness, there is a universal constant  $\pi_{n}$, where $n$ is the number of elements in $\hat{S}(t)$ and $S(t)$, with  $\pi_{n} \rightarrow 1$ as  $n \rightarrow \infty $, and constant  $ C(B)$  depending on the Besov norm  such that
\begin{equation}
   \operatorname{Prob}\left\{\left\|f_{L R}\right\|_{B} \leq C(B)\left\|\hat{f}_{H R}\right\|_{B}\right\} \geq \pi_{n}.
\end{equation}
\label{prop:guarantee}
\end{prop} 

\ff{Prop. \ref{prop:guarantee} reveals the important smoothness relationship between
the original noisy signal and their wavelet recovery. With the overwhelming likelihood and in a common smoothness measure: Besov norm, the recovered signal is smoother than the noisy signal. Prop. \ref{prop:guarantee} offers a theoretical guarantee for the denoising effect of Algorithm \ref{alg:wsa}. Considering that the soft-AE performs a network-based wavelet transform, it inherits such a theoretical guarantee from the wavelet shrinkage theory.}

	\subsection{Gamma Correction}
	To further improve the enhancement effect, the gamma correction can be used for illumination adjustment after obtaining the final reflectance $R$ and illumination $L$. Then, the final result can be obtained by
	\begin{equation}
	S^{\prime}=R\cdot L^{\prime},
	\label{gamma}
	\end{equation}
	where $\cdot$ denotes the element-wise multiplication, and $L^{\prime}=L^{\frac{1}{\gamma}}$. Usually, the value of $\gamma$ is empirically set to 2.2.
	
	This gamma correction method assumes that the illumination map from the Retinex model is close to the real illumination. However, this assumption may not hold for the extremely low illumination. Thus, we adopt the following gamma correction strategy for the extreme low-light images:
	\begin{equation}
	S^{\prime}=R^{\prime}\cdot L^{\prime},
	\label{gamma}
	\end{equation}
	where $R^{\prime}=R^{\frac{1}{\gamma_1}}$, $L^{\prime}=L^{\frac{1}{\gamma_2}}$. We can manually adjust the values of $\gamma_1$ and $\gamma_2$ according to the characteristics of the images. Finally, the whole procedure of the proposed framework is summarized in Algorithm \ref{al}.

\section{Interpretability of the Proposed Framework} 

\ff{The employment of the plug-and-play framework features modularization, which sequentially solves $R$ and $L$ and avoids the crosstalk between them. Our highlight is the plug-and-play framework can enhance interpretability. Generally, interpretability means to what extent a human can understand and reason a model. In \cite{fan2021interpretability}, interpretability is divided into post hoc interpretation and ad hoc interpretable modeling. The former is conducted after a model is well learned. The main advantage of post hoc methods is that one does not need to compromise interpretability with the predictive performance, since prediction and interpretation are separate. The latter is to prototype an interpretable model. The merit of ad hoc interpretable modeling is that it can avoid the bias of post hoc interpretation. We argue that the modularization in the plug-and-play framework can enhance both post hoc and ad hoc interpretability. On the one hand, we can use a powerful denoising network, conjugated by a dedicated post hoc analysis, to simultaneously enjoy satisfactory denoising performance and interpretability. On the other hand, we can straightly utilize an interpretable denoiser. In the following, we illustrate them in detail. } 

\subsection{Post-hoc Analysis}
\ff{Since our framework is essentially iterative, we develop an explanatory directed graph to illustrate how $R$ is solved as the iteration goes. For example, is it local? When and where the information is lost severely? Specifically, inspired by the study in \cite{wang2022ctformer}, we set a pixel $R_i^{t}$ of interest obtained in the $t$-th iteration to zero (mask it) and examine how $R$ in the subsequent iteration is altered. Suppose $R_j^{t+1}$, where $j\neq i$, has a dramatic change, it concludes that $R_i^{t}$ has a major impact on $R_j^{t+1}$. Then, we build a link between them. Repeating this procedure for different pixels and iterative steps, we expect to track the information evolution in the iteration.}

\subsection{Ad-hoc Modeling}

\ff{The aforementioned Soft-AE is essentially unrolling the cascade wavelet shrinkage algorithm into a network, which is naturally more interpretable than the conventional autoencoders. Let us mathematically illustrate the connection between the Soft-AE and the wavelet shrinkage system.}

\ff{Without loss of generality, we consider a four-convolutional-layer Soft-AE. Suppose that this four-convolutional-layer Soft-AE contains $N$  filters in the first encoding layer and  $M\times N$  filters in the second encoding layer, which are denoted as  $\psi_{i}, i\in [N]$ and $\psi_{i j}, i\in [M]$; $j\in [N]$, respectively. In symmetry, the two decoding layers of this four-convolutional-layer Soft-AE consists of $N\times M$  and $N$ filters, which are denoted as $\phi_{i j}, i\in [N]$; $j\in [M]$ and $\phi_{i}, i\in [N]$, respectively. For simplicity, we use $(\cdot)^+$ to denote the soft thresholding function $\eta_b(\cdot)$. The final output of this four-convolutional-layer Soft-AE is
\begin{equation}
    \sum_{k}^{N} \phi_{k} *\left[\sum_{j}^{M} \phi_{k j} *\left[\sum_{i}^{N} \psi_{j i} *\left(\psi_{i} *x\right)^{+}\right]^{+}\right],
    \label{eqn:output}
\end{equation}
where $*$ represents convolution. We can apply the approximate property of the soft thresholding:
\begin{equation}
   (h+g)^{+} \sim h^{+}+g^{+}. 
   \label{eqn:soft-thresholding}
\end{equation}
When the threshold is zero, a soft-thresholding activation degenerates into a linear activation. Thus, Eq. \eqref{eqn:soft-thresholding} approximately holds when the threshold is small. As a result, Eq. \eqref{eqn:output} turns into
\begin{equation}
\sum_{k}^{N} \phi_{k} * \sum_{j}^{M}\left[\phi_{k j} * \sum_{i}^{N} \psi_{j i} *\left(\psi_{i} * x\right)^{+}\right]^{+} .
\label{eqn:output_simple}
\end{equation}}

\ff{Let $\Psi $ be a matrix of the size of $M \times N$ whose $(j,i)$-entry is  $\psi_{j i}$, and  $\Phi$ be the  $N \times M$  matrix whose $(k,j)$-entry is $\phi_{k j}$. Since convolution operations conform to the associative laws, Eq. \eqref{eqn:output_simple} is further simplified into the matrix form:
\begin{equation}
    \left[\phi_{1}, \ldots, \phi_{N}\right] \otimes \Phi \otimes \Psi \otimes\left[\left(\psi_{1} * x\right)^{+}, . .,\left(\psi_{N} * x\right)^{+}\right]^{T},
\end{equation}
where  $(A \otimes B)_{i j}=\sum_{k} A_{i k} * B_{k j}$, which is analogous to the matrix product but elements are convolutional filters, and the operation between the elements is convolution. The Soft-AE is to do wavelet shrinkage, and can recover the clean signal, when the following conditions are satisfied:
\begin{equation}
    \left\{\begin{array}{c}
\Phi \otimes \Psi=\operatorname{diag}\left(\lambda_{1}, \lambda_{2}, \ldots, \lambda_{N}\right) \delta \\
\phi_{k}=\frac{\psi_{k}^{-1}}{\left|\sum_{k}^{N} \lambda_{k}\right|} \text { or } \psi_{k}=\frac{\phi_{k}^{-1}}{\left|\sum_{k}^{N} \lambda_{k}\right|}, k=1,2, \ldots, N
\end{array},\right.
\label{eqn:condition}
\end{equation}
where $\delta$ is the Dirac function, and the selection of  $\Phi \otimes \Psi $ should make $\sum_{k}^{N} \lambda_{k} $ nonzero. Eq. \eqref{eqn:condition} can be trivially fulfilled by setting diagonal elements of  $\Phi$  and  $\Psi$  to be mutually inverse to each other and the rest elements to zero.}

	\begin{figure*}[b!]
		\centering
		\subfloat[Input]{   \includegraphics[width=0.16\textwidth]{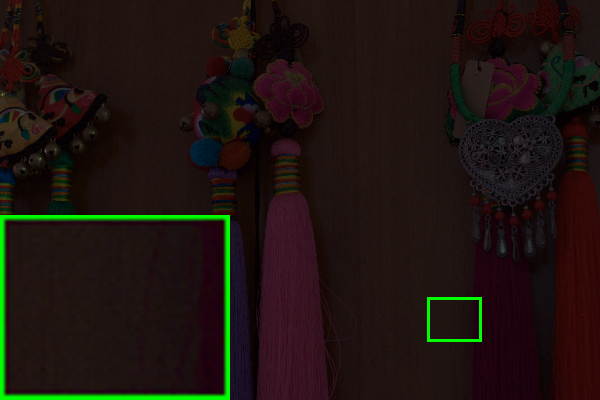}}
		\subfloat[LIME\cite{guo2016lime}]{	\includegraphics[width=0.16\textwidth]{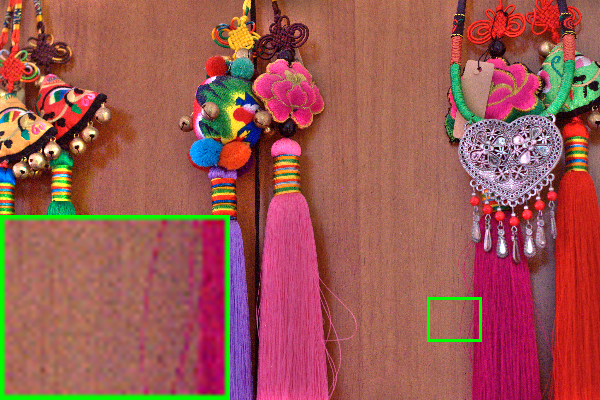}}		
		\subfloat[RetinexNet\cite{wei2018deep}]{	\includegraphics[width=0.16\textwidth]{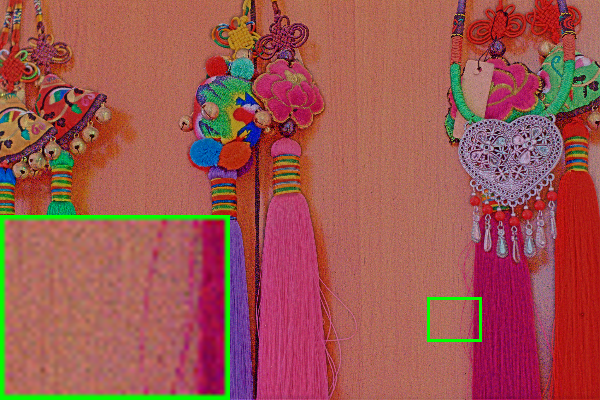}}  
		\subfloat[Zero-DCE\cite{guo2020zero}]{	\includegraphics[width=0.16\textwidth]{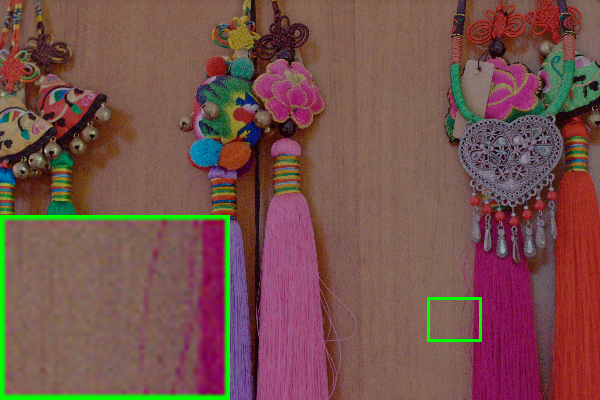}}	
		\subfloat[KinD\cite{zhang2019kindling}]{	\includegraphics[width=0.16\textwidth]{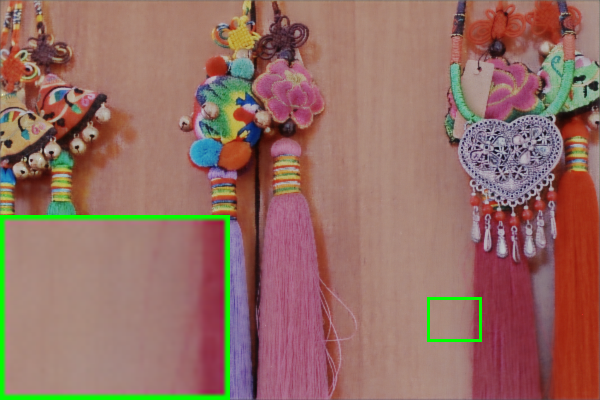}}	
		\subfloat[FOTV\cite{gu2019novel}]{	\includegraphics[width=0.16\textwidth]{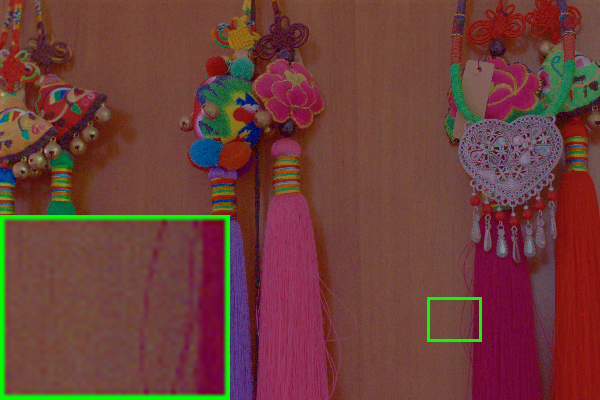}}
		
		\subfloat[LR3M\cite{ren2020lr3m}]{	\includegraphics[width=0.16\textwidth]{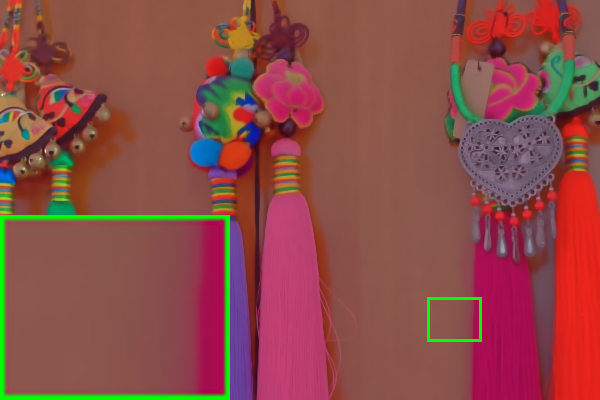}}		
		\subfloat[\ff{RetinexDIP\cite{zhao2021retinexdip}}]{	\includegraphics[width=0.16\textwidth]{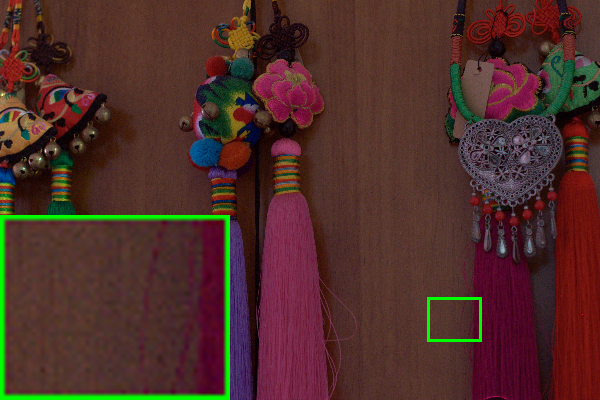}}
		\subfloat[\ff{URetinex\cite{wu2022uretinex}}]{	\includegraphics[width=0.16\textwidth]{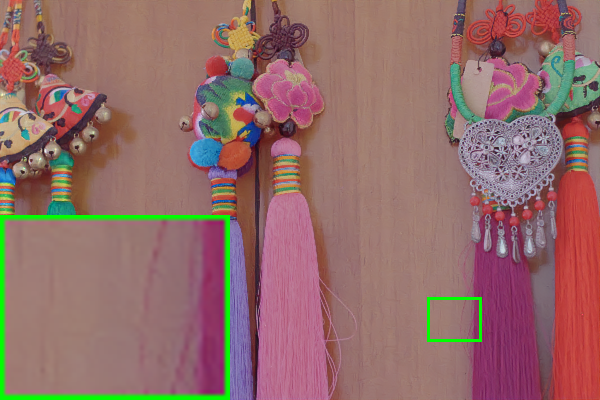}}	
		\subfloat[Ours-IRCNN]{	\includegraphics[width=0.16\textwidth]{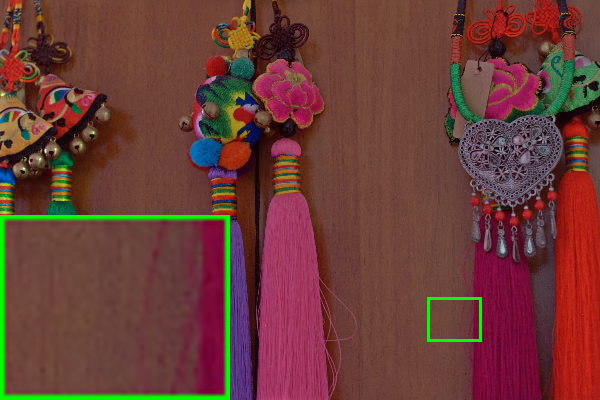}}	
		\subfloat[Ours]{	\includegraphics[width=0.16\textwidth]{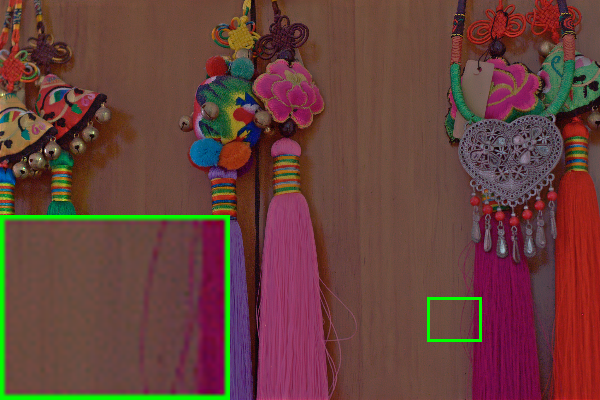}}	
  \subfloat[Ground Turth]{	\includegraphics[width=0.16\textwidth]{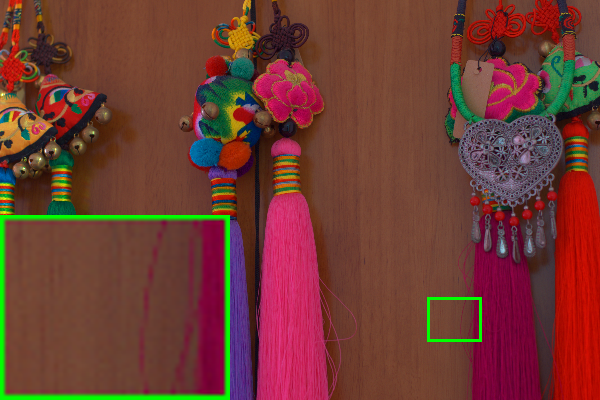}}

		\caption{The enhanced results of existing methods on LOL179.}
		\label{our1}
	\end{figure*}
	
	\begin{figure*}[b!]
		\centering
		\subfloat[Input]{   \includegraphics[width=0.16\textwidth]{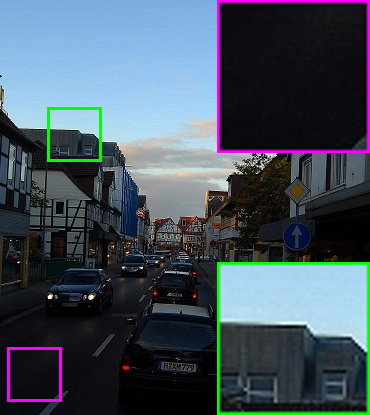}} 
		\subfloat[LIME\cite{guo2016lime}]{	\includegraphics[width=0.16\textwidth]{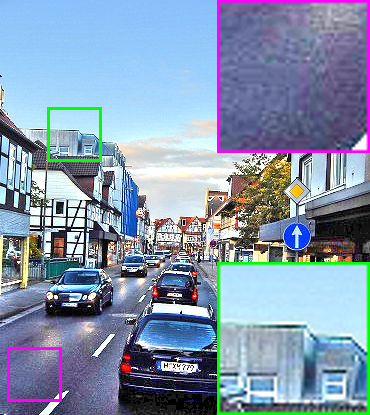}}	
		\subfloat[RetinexNet\cite{wei2018deep}]{	\includegraphics[width=0.16\textwidth]{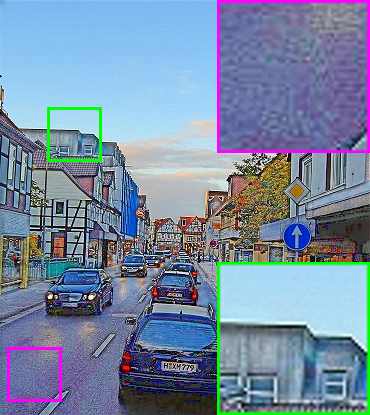}}
		\subfloat[Zero-DCE\cite{guo2020zero}]{	\includegraphics[width=0.16\textwidth]{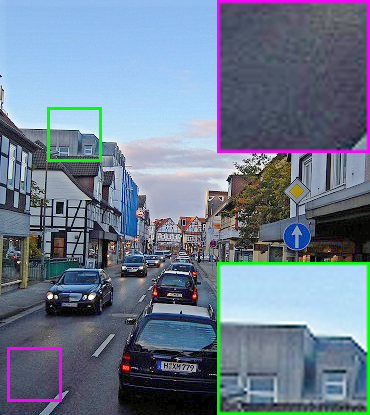}}
		\subfloat[KinD\cite{zhang2019kindling}]{	\includegraphics[width=0.16\textwidth]{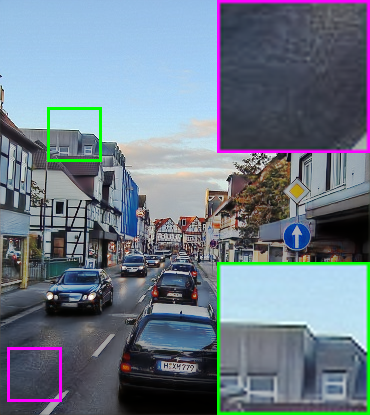}}	
		\subfloat[FOTV\cite{gu2019novel}]{	\includegraphics[width=0.16\textwidth]{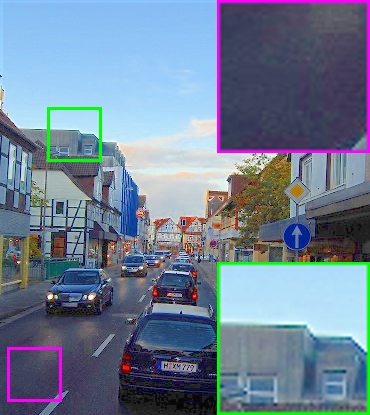}}
		
		\subfloat[LR3M\cite{ren2020lr3m}]{	\includegraphics[width=0.16\textwidth]{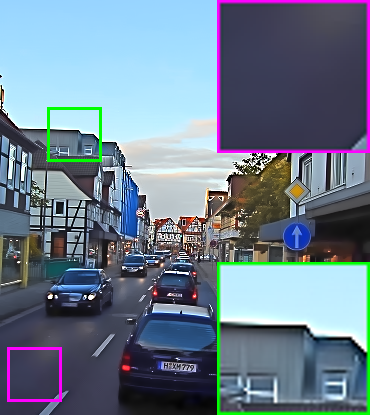}}	
		\subfloat[\ff{RUAS\cite{liu2021retinex}}]{	\includegraphics[width=0.16\textwidth]{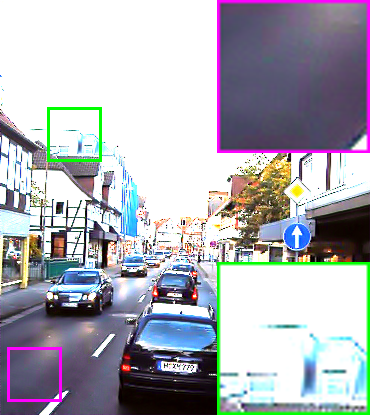}}  
		\subfloat[\ff{RetinexDIP\cite{zhao2021retinexdip}}]{	\includegraphics[width=0.16\textwidth]{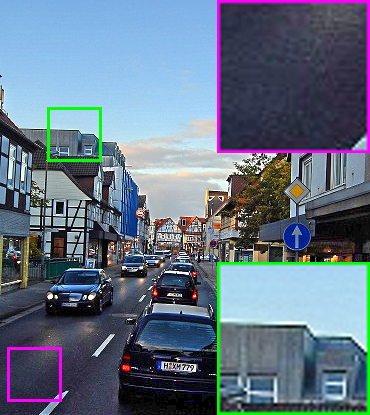}}
		\subfloat[\ff{URetinex\cite{wu2022uretinex}}]{	\includegraphics[width=0.16\textwidth]{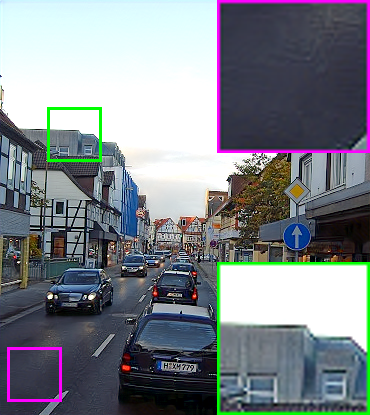}}	
		\subfloat[Ours-IRCNN]{	\includegraphics[width=0.16\textwidth]{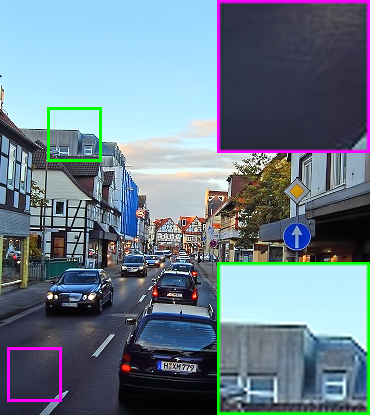}}
		\subfloat[Ours]{	\includegraphics[width=0.16\textwidth]{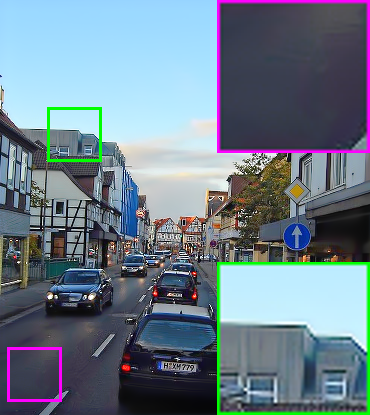}}	
		\caption{The image enhancement results of existing methods on Img7.}
		\label{our2}
	\end{figure*}

	\section{Main Experiments} \label{experiment}

	\ff{Extensive experiments are conducted to evaluate the effectiveness of our framework. First, we introduce the experimental protocols. Second, we analyze via qualitative visualization and quantitative comparisons the image enhancement performance of our proposed framework, with an emphasis on the effects of the initial illumination map and the proposed sequential decomposition.} 

	\subsection{Experimental Protocols}
	\subsubsection{Datasets} 
\begin{figure}[h]
		\centering
		\includegraphics[scale=0.28]{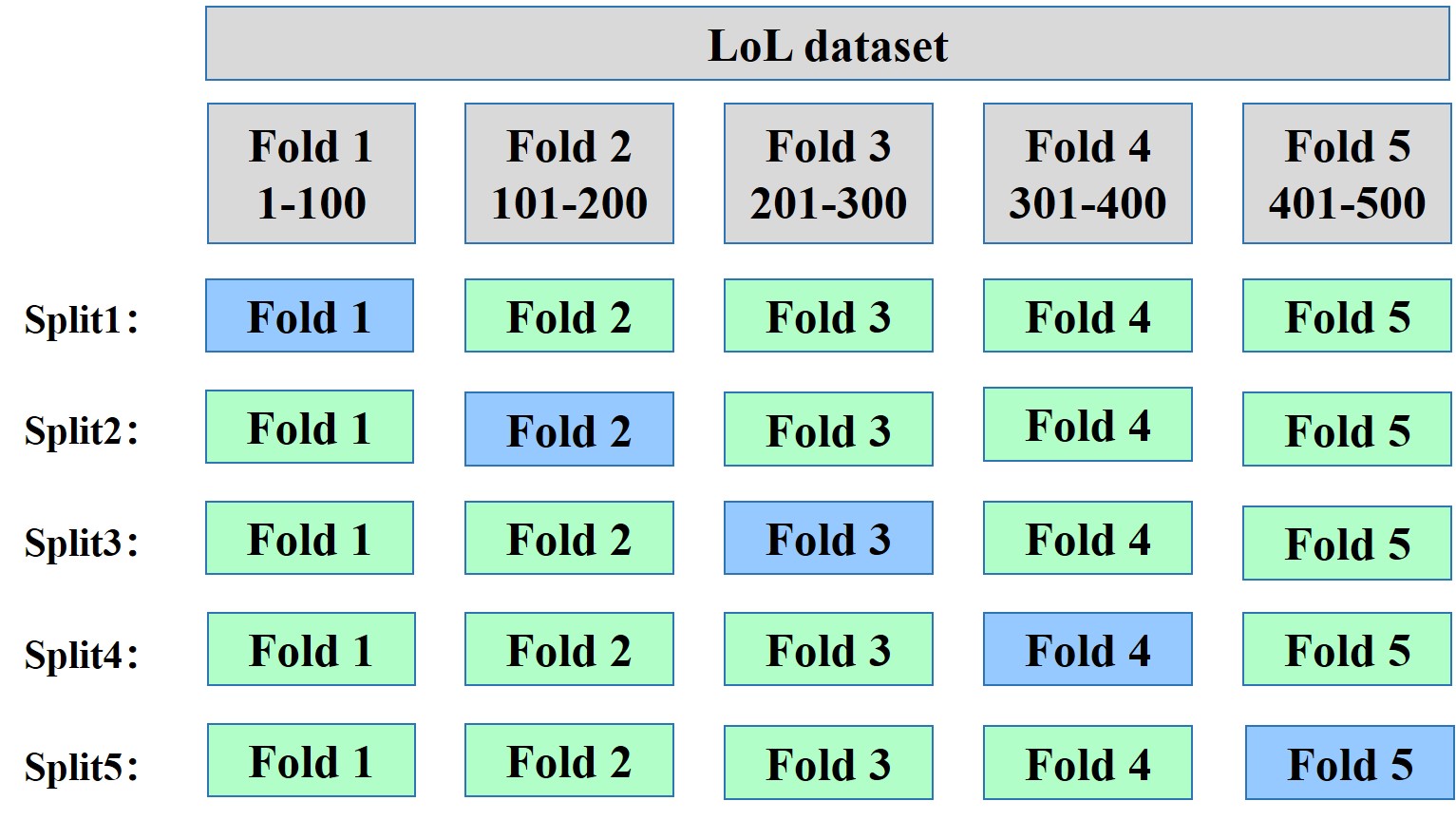}
		\caption{\ff{The split of the 5-fold cross-validation. Especially, the LOL dataset contains 500 images, and these images are randomly coded from 1 to 500.}}
		\label{5fold}
	\end{figure}

We conduct extensive experiments on three datasets. The first dataset is Set12, which contains 12 underexposed images with real noise collected in public datasets\cite{guo2016lime,lee2012contrast,wang2013naturalness}.
	The second dataset is LOL \cite{wei2018deep} (\url{https://daooshee.github.io/BMVC2018website/}) which contains 500 low-light images and their normal-light ground truth. We use the 5-fold cross-validation to test the performance of all supervised learning-based methods on the LOL dataset.
	Fig. \ref{5fold} showcases the splits of the 5-fold cross-validation. The third dataset is Berkeley segmentation dataset (BSD) (\url{https://www2.eecs.berkeley.edu/Research/Projects/CS/vision /bsds/}) that contains 30 images with normal light and is employed for testing. We use a simple synthetic strategy to transform those normal-light images into low-light versions with noise distortion. First, we generate the low-light image by reducing the brightness of the V-channel of the normal images in HSV color space. Then, the white Gaussian noise with $\sigma=5$ is added into the low-light images to obtain the desired images. 

	\subsubsection{Compared Methods} To demonstrate the superiority of our method, we compare it with 9 off-the-shelf advanced image enhancement approaches, including 3 traditional methods FOTV\cite{gu2019novel}, LIME\cite{guo2016lime} and LR3M\cite{ren2020lr3m}, and 6 machine learning-based methods RetinexNet\cite{wei2018deep},  KinD\cite{zhang2019kindling}, Zero-DCE\cite{guo2020zero}, RetinexDIP\cite{zhao2021retinexdip}, RUAS\cite{liu2021retinex}, and URetinex\cite{wu2022uretinex}. \ff{All these methods are either classical benchmarks or state-of-the-arts published in recent two years.} We configure the parameters of all compared methods based on the recommendation of the original papers.
	
	\subsubsection{Evaluation Metric} Two types of image quality assessment indexes are employed. The first type is full-reference image quality assessment metrics (FR-IQAs): PSNR, SSIM, and MSE which are used when both low-light images and normal-light ground truth are available. The second type is no-reference image quality assessment metrics (NR-IQAs) which are suitable when only low-light images are available. The widely-used NR-IQAs are NIQE\cite{mittal2012making}, BTMQI\cite{gu2016blind}, and ARISMC\cite{gu2015no}. The lower NIQE, BTMQI, and ARISMC, the better perceptual quality. 
	

	\subsection{Performance Analysis}
	\subsubsection{Qualitative Comparison}
	Figs. ~\ref{our1} and \ref{our2} exhibit the visual comparison between the proposed method and competitors on LOL179 and Img7.
	\begin{table}
		\centering
		\caption{Average and std numerical results on Set12. The best performer is bold-faced. The runner-up is underlined.}
		\label{table1}
		\setlength{\tabcolsep}{0.8mm}{
			\begin{tabular}{p{55pt}cccc}  
				\toprule                              
	Method&NIQE&BTMQI&ARISMC&Rank\\\midrule 
				LIME\cite{guo2016lime}&3.1546$\pm$0.8511&4.6255$\pm$0.8585&3.4193$\pm$0.9463&10\\
				FOTV\cite{gu2019novel}&3.0282$\pm$0.7282&3.6458$\pm$0.8939&3.4218$\pm$0.8790&6\\
				LR3M\cite{ren2020lr3m}&3.4828$\pm$0.6605&3.8042$\pm$1.2977&2.6589$\pm$0.3321&7\\
				RetinexNet\cite{wei2018deep}&4.6671$\pm$0.9783&3.4078$\pm$0.8609&3.0486$\pm$0.5056&7\\
				KinD\cite{zhang2019kindling}&3.0295$\pm$0.6271&3.7100$\pm$1.4252&\textbf{2.5341$\pm$0.1184}&3\\
				Zero-DCE\cite{guo2020zero}&2.8006$\pm$0.8096&3.5091$\pm$1.1919&2.7711$\pm$0.2568&4\\
				\ff{RUAS\cite{liu2021retinex}}&\ff{4.0258$\pm$1.3475}&\ff{5.2329$\pm$1.7881}&\ff{2.9608$\pm$0.4262}&11\\
                \ff{RetinexDIP\cite{zhao2021retinexdip}}&\ff{3.4750$\pm$0.7543}&\ff{4.2005$\pm$1.2592}&\ff{3.1618$\pm$0.5048}&9\\			
				\ff{URetinex\cite{wu2022uretinex}}&\ff{3.4508$\pm$0.8024}&\ff{3.4446$\pm$1.3577}&\ff{2.6506$\pm$0.1011}&5\\	
				\ff{Ous-IRCNN}&\ff{\underline{2.7776$\pm$0.7051}}&\ff{ \underline{3.3741$\pm$1.4002}}&\ff{2.6000$\pm$0.1222}&2\\	Ours&\textbf{2.7198$\pm$0.7934}&\textbf{3.3436$\pm$1.0732}&\underline{2.5780$\pm$0.1072}&\textbf{1}\\\bottomrule
		\end{tabular}}
	\end{table}	

 	\begin{table*}[b]\tiny
		\centering
		\caption{Average and std numerical results on LOL dataset. The best performer is bold-faced. The runner-up is underlined.}
		\setlength{\tabcolsep}{3pt}
  \scalebox{1.3}{
		\begin{tabular}{lccccccccc}    
			\toprule                             
			\textbf{Average$\pm$std}&FOTV\cite{gu2019novel}& LIME\cite{guo2016lime}&LR3M\cite{ren2020lr3m}&RetinexNet\cite{wei2018deep}&Zero-DCE\cite{guo2020zero}&KinD\cite{zhang2019kindling}&\ff{RetinexDIP\cite{zhao2021retinexdip}}&\ff{RUAS\cite{liu2021retinex}}&Ours \\\midrule 
			PSNR&13.70$\pm$4.14&15.24$\pm$2.59&14.67$\pm$4.24&15.39$\pm$3.22&14.16$\pm$4.71&\underline{15.98$\pm$4.81}&8.94$\pm$3.20&14.24$\pm$4.29&\textbf{17.12}$\pm$\textbf{3.14}\\ 
			SSIM&0.5155$\pm$0.1703&0.4460$\pm$0.1556&0.6135$\pm$0.1775&0.5461$\pm$0.1594&0.5318$\pm$0.1893&\underline{0.6578$\pm$0.1803}&0.2949$\pm$0.2022&0.4786$\pm$0.1670&\textbf{0.6651$\pm$0.1667}\\ 
		    MSE&4019.4$\pm$3247.6&\underline{2347.1$\pm$1634.8}&3301.4$\pm$2824.4&2474.7$\pm$2137.7&3962.5$\pm$3627.4&2828.1$\pm$3044.1&10286.0$\pm$5957.3&3982.5$\pm$4381.8&\textbf{1667.3}$\pm$\textbf{1465.6}\\
			NIQE&8.9681$\pm$1.7472&9.3741$\pm$1.9435&4.6112$\pm$0.7123&6.8250$\pm$1.3304&8.5394$\pm$1.8042&\underline{4.3496$\pm$1.1772}&6.9270$\pm$1.3364&6.1581$\pm$1.3799&\textbf{3.7051}$\pm$\textbf{0.8058}\\	
            BTMQI&4.9480$\pm$1.1103&\underline{3.7655$\pm$0.8957}&5.5673$\pm$1.2495&\textbf{3.1050$\pm$1.2812}&4.6587$\pm$1.3103&4.5545$\pm$1.1164&6.2707$\pm$1.0134&4.6972$\pm$1.5844&4.4144$\pm$1.6022\\
			ARISMC&3.3168$\pm$0.3947&3.4268$\pm$0.4454&\textbf{2.4946$\pm$0.2249}&3.1503$\pm$0.1411&3.3750$\pm$0.3344&2.9158$\pm$0.2324&3.2709$\pm$0.2415&3.0668$\pm$0.1517&\underline{2.7255$\pm$0.2121}\\
			Ave. Rank&8&6&4&3&7&2&9&5&\textbf{1}
			\\\bottomrule
		\end{tabular}}
		\label{table2}
	\end{table*}
	The highlights of Fig. \ref{our1} are as follows.
    First, the RetinxDIP method only shows moderate enhancement effects. Most regions of this enhanced image are still extremely dark and insufficiently contrastive.
    Second, although LIME, RetinexNet, Zero-DCE, and FOTV obtain the desirable enhancement, they fail to eliminate the noise in the dark regions.
    Third, as shown in the zoomed regions, the visual results generated by KinD and LR3M are over-smooth, resulting in unclear and unsharp structural expression, while the URetinex and KinD produce unrealistic results with color distortion.
	Last, the proposed model is visually superior, with a better enhancement and a lower noise level. The above-mentioned issues such as heavy noise appearing in the dark regions and unclear structures are markedly alleviated.
	
	From Fig. \ref{our2}, we draw four observations.
	First, most methods perform poorly in denoising, \textit{e.g.}, significant noise remains on the road. Although LR3M and RUAS effectively suppress noise, the texture in the green box is missing because of the over-denoising effect. 
	Second, the results of LIME, RetinexNet, and RUAS are over-enhanced, \textit{e.g.}, extra color distortion is present.
	Third, URetinex and RUAS unsatisfactorily overexpose the sky.
	Last, the visual effect created by our proposed model is pleasing. For example, the structures are sharper, and the surface is clearer.

  	\begin{figure*}[t]
		\centering
		\subfloat[]{   \includegraphics[height=0.12\textwidth]{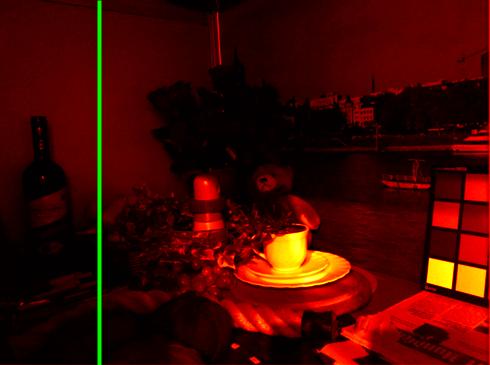}} 
		\subfloat[]{   \includegraphics[height=0.12\textwidth]{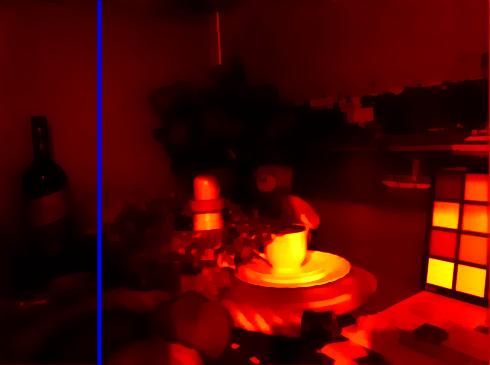}} 
		\subfloat[]{	\includegraphics[height=0.12\textwidth]{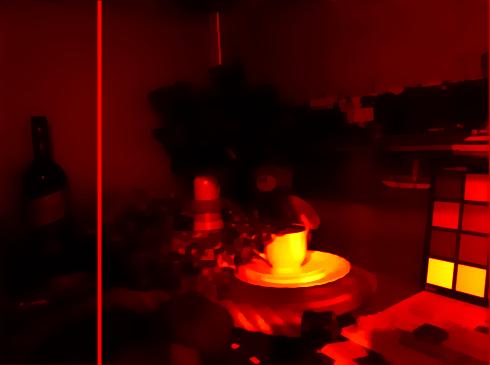}}
		\subfloat[]{	\includegraphics[height=0.12\textwidth]{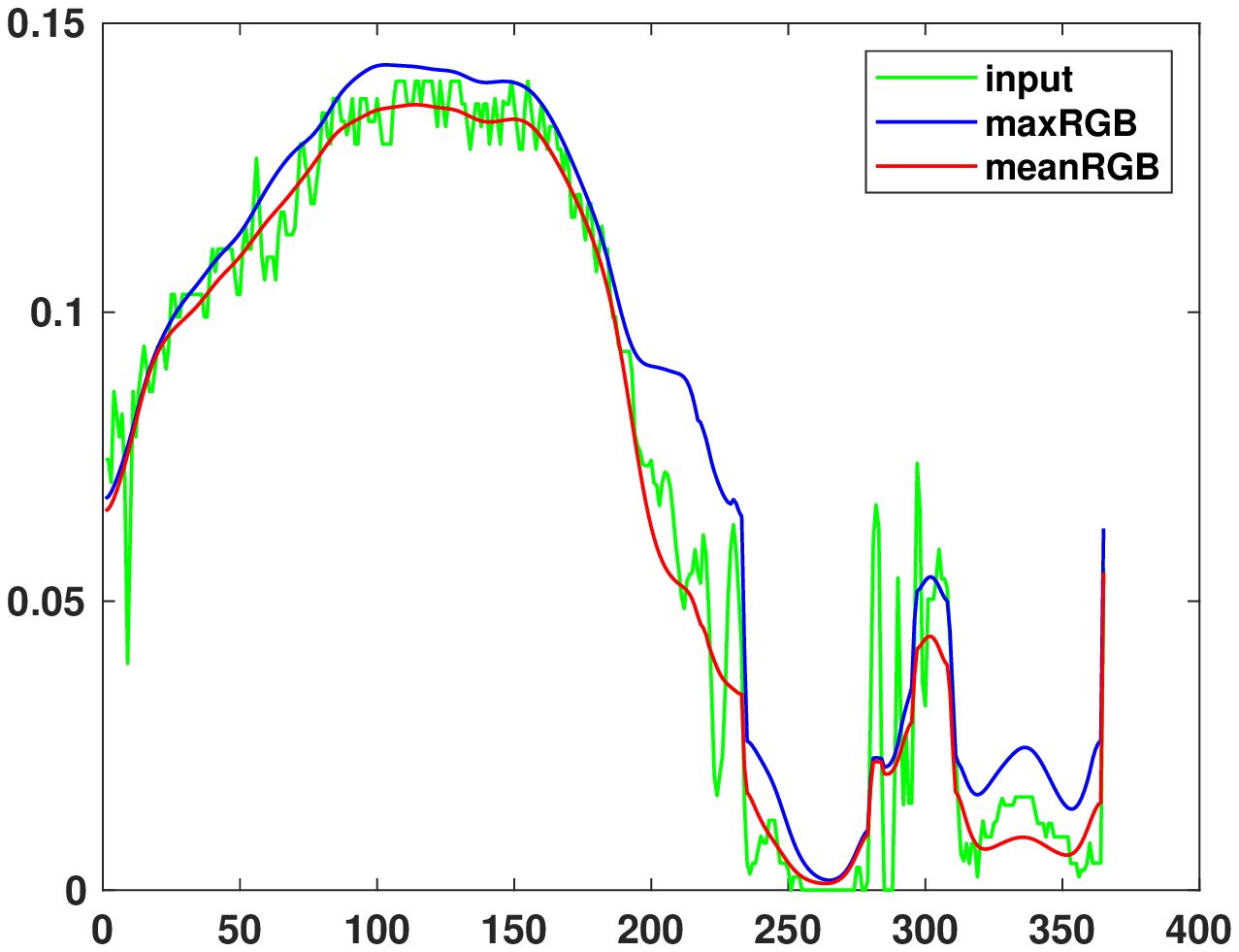}}
		\subfloat[NIQE: 3.3889]{	\includegraphics[height=0.12\textwidth]{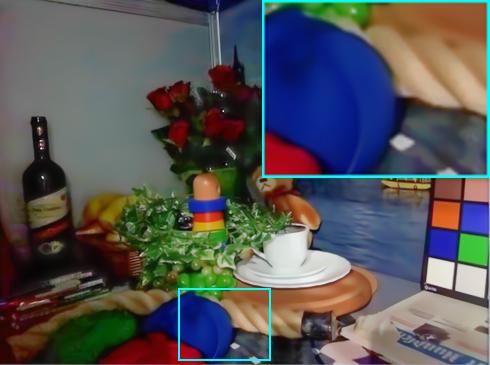}}
		\subfloat[NIQE: 3.1811]{	\includegraphics[height=0.12\textwidth]{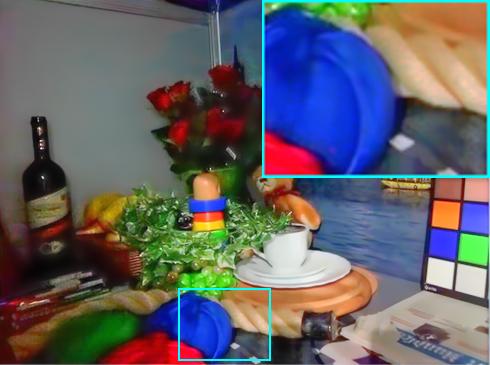}}
		\caption{\ff{Comparison between the meanRGB operator and maxRGB operator. Heatmap of (a) input, (b) maxRGB and (c) meanRGB. (d) 1-D plots of the estimated $L$ map. Enhanced results of (e) maxRGB and (f) meanRGB.}}
		\label{rgb}
	\end{figure*}
	
	\subsubsection{Quantitative Comparison}
	Now we quantitatively compare our framework with its competitors on two test datasets. Specifically, we use the NR-IQAs to measure the performance of all the methods on Set12, and the NR-IQAs and FR-IQAs on the LOL dataset.
	
	Tabs. \ref{table1} and \ref{table2} summarize the experimental results of all methods on LOL and Set12, respectively. Note that the Ave. Rank is the comprehensive ranking of all numerical results. 
	First, the regularization-based algorithms show suboptimal performance on the Set12 and LOL because those methods ignore the noise distortion in enhancing images.
	Second, compared to the regularization-based methods, learning-based models achieve better results. Since the performance of pure learning-based methods relies heavily on the number of training samples, and this dataset only has limited data, their results are not completely satisfactory.
	Last, among all the image enhancement models, our method obtains the best numerical scores in light of all metrics. 
	In addition, our model has a relatively low std in the LOL dataset, suggesting that our model is more robust and consistent.

	\begin{table}
		\centering
		\caption{Average numerical results on 30 synthetic low-light images with noise distortion}
		\label{nstab}
		\setlength{\tabcolsep}{3pt}
		\scalebox{1.0}{
			\begin{tabular}{p{50pt}cccc}  
				\toprule                     	
				Method&PSNR&SSIM&MSE\\\midrule  
				RetinexNet\cite{wei2018deep}&14.03$\pm$2.14&0.7059$\pm$0.1242&2870.4$\pm$1368.3\\
				KinD\cite{zhang2019kindling}&15.33$\pm$2.39&0.7581$\pm$0.1005&2118.3$\pm$1132.8\\
				Zero-DCE\cite{guo2020zero}&17.39$\pm$1.79&0.7576$\pm$0.0850&1278.6$\pm$493.9\\
				\ff{RUAS\cite{liu2021retinex}}&\ff{15.09$\pm$1.69}&\ff{0.7267$\pm$0.0870}&\ff{2155.2$\pm$788.5}\\
				\ff{URetinex\cite{wu2022uretinex}}&\ff{16.07$\pm$3.18}&\ff{0.7605$\pm$0.1096}&\ff{2025.6$\pm$1415.4}\\			Ours&\textbf{19.89$\pm$2.18}&\textbf{0.8328}$\pm$\textbf{0.0678}&\textbf{741.7}$\pm$\textbf{327.2}\\\bottomrule 
		\end{tabular}}
	\end{table}
 
 	To further test the effectiveness of the proposed method in noise suppression, here we specifically compare it with five competitors that also consider noise suppression in image enhancement. The images are from the BSD dataset. Table \ref{nstab} reports the quantitative results on the synthetic 30 low-light and noisy images. In Table \ref{nstab}, our method achieves the highest PSNR and SSIM values and the lowest MSE error.

	\subsubsection{\ff{The initial illumination map}}
	\ff{Different from the methods based on alternating iteration, our model estimates $R$ and $L$ separately. The key to the success of the proposed framework is to accurately estimate $L$.
		In addition to the meanRGB operator mentioned in this paper, the maxRGB operator is often used to estimate the initial illumination map, which is mathematically defined as}
	\begin{equation}
	\ff{
		\hat{L}(x)=\max_{c\in\{R,G,B\}}L^c(x).}
	\end{equation}
	\ff{We compare the performance of two initialization operators in Fig. \ref{rgb}, where
		(a)-(c) are heatmaps of the input image and the estimated illumination obtained by maxRGB, meanRGB.
		As can be seen, both methods are noise-free, but the estimated illumination obtained by meanRGB is smoother than that obtained by maxRGB.
		The 1D profile in Fig. \ref{rgb}(d) also implicates that the pixel profile of the illumination map obtained by meanRGB is more faithful to the input image. Furthermore, we can see in Fig. \ref{rgb}(e)-(f) that the meanRGB initial illumination yields a better enhancement result.
	}

	\subsubsection{Retinex decomposition}
\begin{figure}
		\centering	
		\subfloat[]{
			\includegraphics[width=0.11\textwidth]{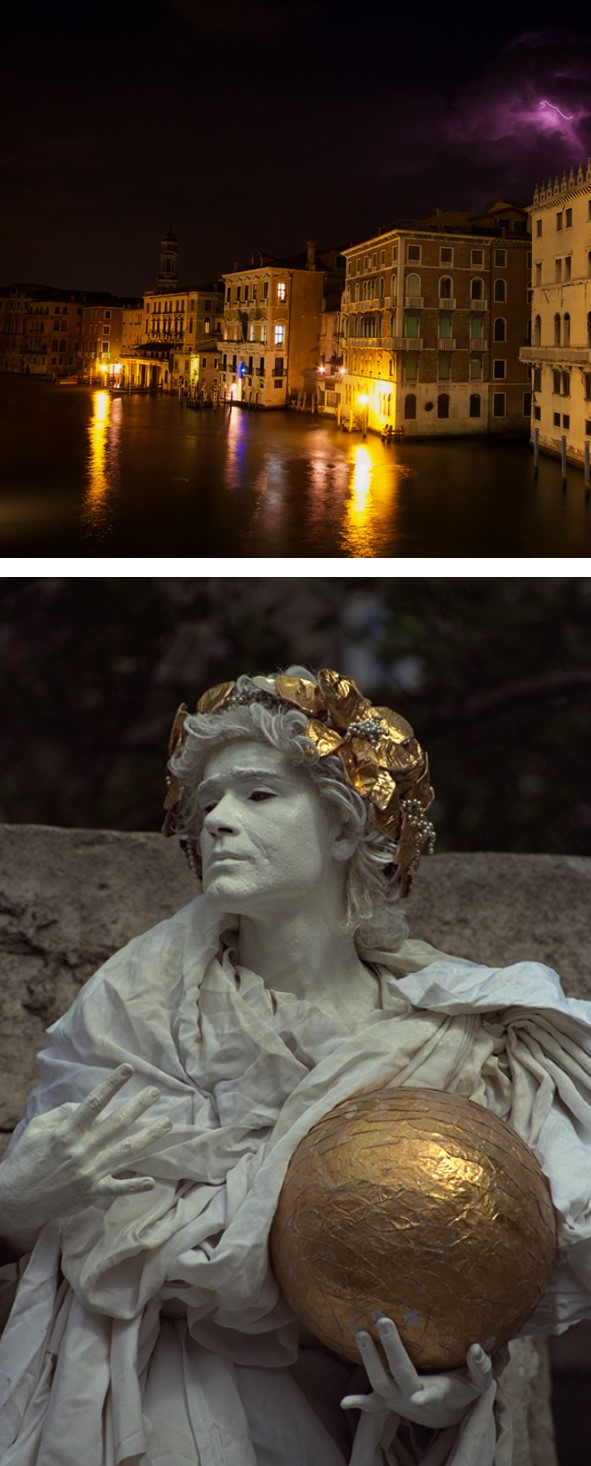}}
		\subfloat[]{
			\includegraphics[width=0.11\textwidth]{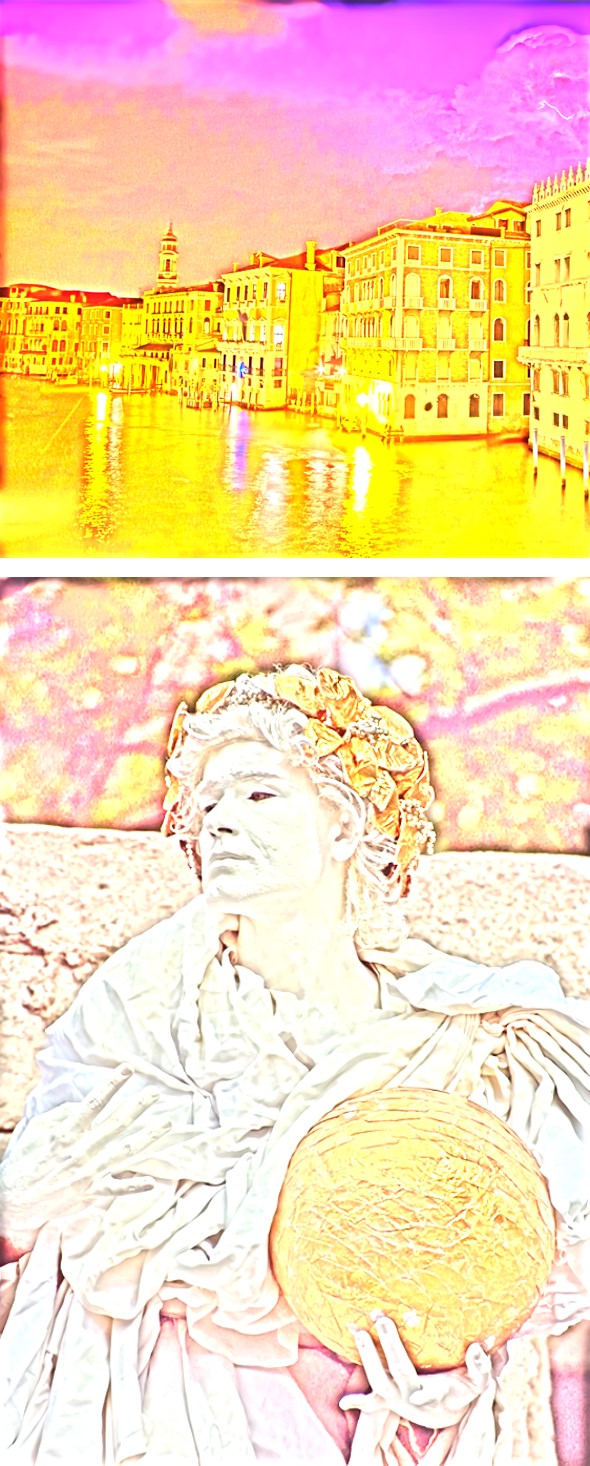}\label{Rk}}	
		\subfloat[]{
			\includegraphics[width=0.11\textwidth]{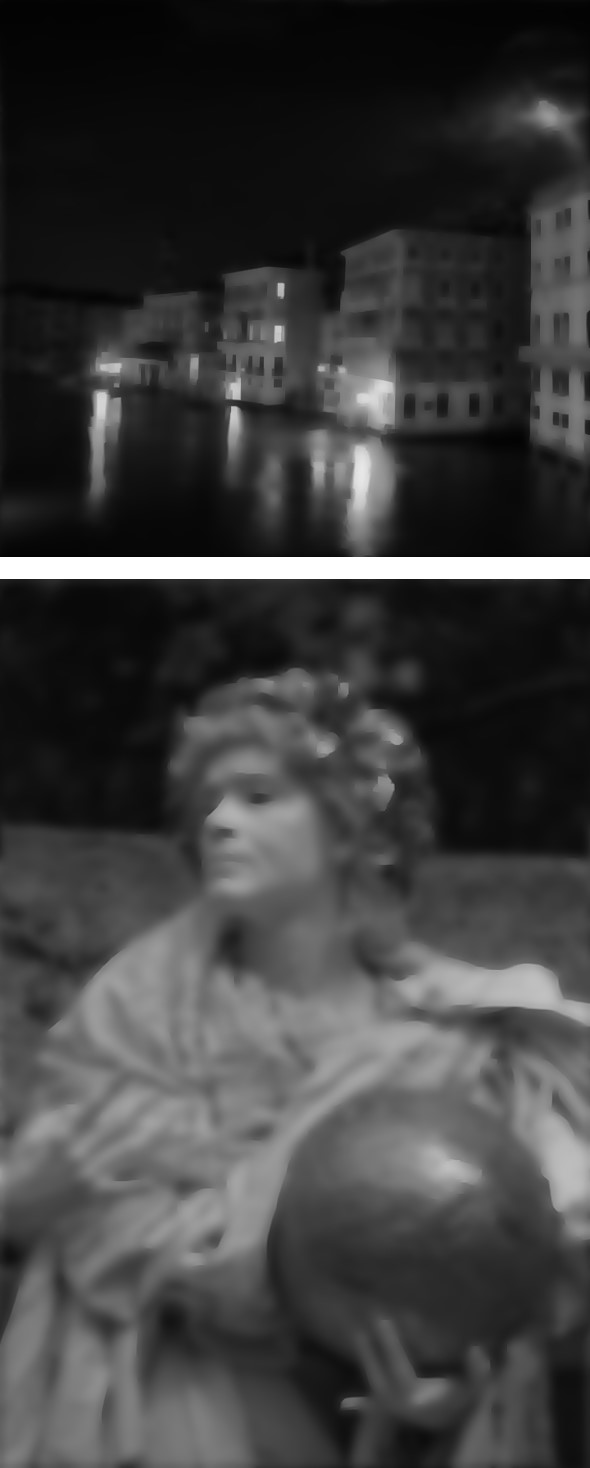}\label{decomb}}	
		\subfloat[]{
			\includegraphics[width=0.11\textwidth]{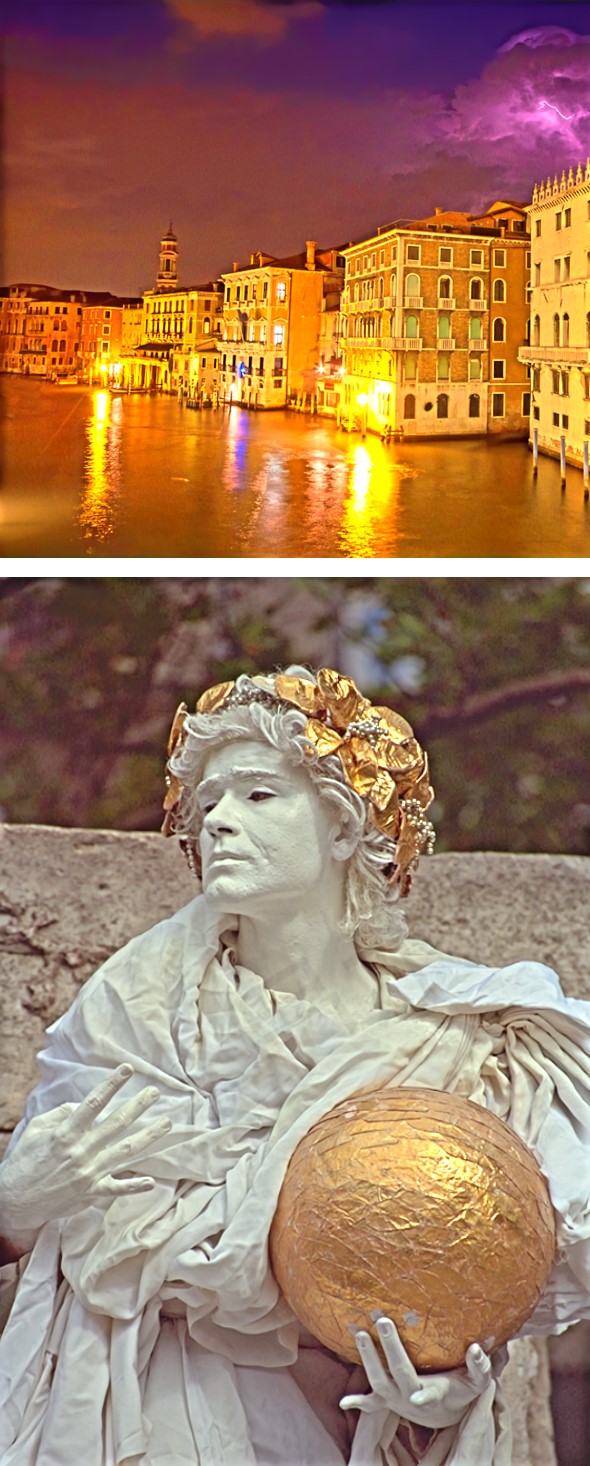}}
		\caption{The decomposition results generated by our framework. (a), (b), (c) and (d) represent the observed image, the estimated reflectance $R^{\left(k\right)}$, illumination $L$, and enhanced version, respectively. }
		\label{decom}
	\end{figure}

	To validate the effectiveness of our decomposition strategy, we offer the decomposition results of our framework in Fig. \ref{decom}, where (a)-(d) represent the input $S$, the reflectance layer $R^{(k)}$, illumination layer $L$, and the restored result, respectively.
	In our framework, the noise existing in the illumination component is transferred to the reflectance layer.
	Thus, the illumination layer should be spatially smooth and only contains simple structures. We only need to denoise the reflectance map.
	Fig. \ref{decom}(b) spotlights that our method can effectively remove the noise in the reflectance map and preserve the edge and structure information.
	In addition, the estimated illumination map (Fig. \ref{decom}(c)) is smooth and noise-free.
	Those observations conclude the effectiveness of the decomposition strategy and justify the utility of our plug-and-play framework for image enhancement.

 	\subsubsection{Runtime}
		\ff{Table \ref{runtime} demonstrates the runtime of our method and other state-of-the-arts for a single image. 
Note that the proposed model is faster than LR3M and RetinexDIP but slower than other models.
Although our model is not the fastest, it has moderate runtime. We argue that a higher time cost is worthwhile in order to acquire excellent enhancement performance.  
The most time-consuming part of our framework lies in deriving $L$. The denoising time of our denoiser only takes 0.2601s.}

	\begin{table}
		\centering
		\caption{\ff{Runtime of different models for a single image (Img1).}}
		\label{runtime}
		\setlength{\tabcolsep}{3pt}
		\begin{tabular}{p{80pt}c}  
			\toprule                               
			Method&Runtime\\\midrule 
			LIME\cite{guo2016lime}&4.2861s\\
			FOTV\cite{gu2019novel}&4.7405s\\
			LR3M\cite{ren2020lr3m}&106.6291s\\ \midrule 
			RetinexNet\cite{wei2018deep}&1.8547s\\
			Zero-DCE\cite{guo2020zero}&2.3340s\\
			KinD\cite{zhang2019kindling}&1.6160s\\	RetinexDIP\cite{zhao2021retinexdip}&24.4460s\\
			RUAS\cite{liu2021retinex}&2.3995s\\
			URetinex\cite{wu2022uretinex}&1.1920s\\ \midrule 
			Ours&16.1763s\\\bottomrule
		\end{tabular}
	\end{table}


	\subsection{\ff{Interpretability of the Proposed Framework}}\label{interpretability}

\ff{Earlier experiments show that our designed denoisier can assist the framework to deliver advanced image enhancement performance. Now, we use the aforementioned post hoc dynamics analysis to track the denoising process to address problems such as what kind of information is used in denoising in our framework. }

\begin{figure}[h]
		\centering
		\includegraphics[scale=0.6]{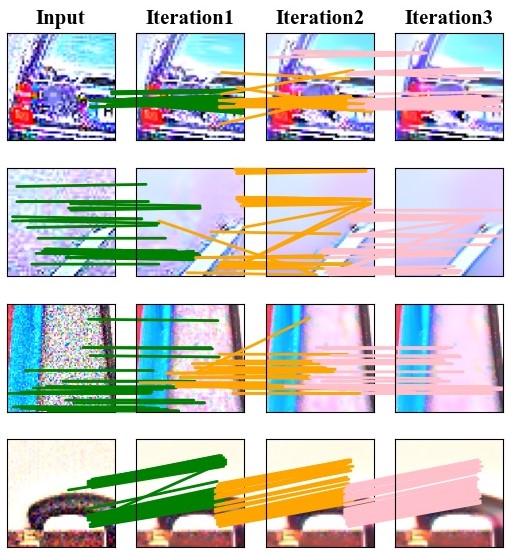}
		\caption{The information flow of the proposed framework.}
		\label{information}
	\end{figure}

\ff{The directed graph is built based on the algorithm in Section IV. A. We demonstrate the information flow of denoising in Fig. ~\ref{information}. Each arrow connects the most related two pixels between images generated by two iterations. From Fig.~\ref{information}, we draw one interesting observation. It can be seen that the information flow of denoising is highly local, \textit{i.e.}, the restoration of some pixels is most influenced by its very surrounding pixels. This is surprising because the deeper layers of a CNN usually can extract abstract information which can cover a large receptive field. Thus, the interaction between two pixels is expected to be local but not highly local. Since various CNN-based denoisers with interpretability can be inserted at will, we conjecture that if we insert a transformer \cite{vaswani2017attention} into the denoising module, we might see the interaction in a much longer range.} 

	\section{Ablation Study and Parametric Analysis}
 \subsection{Ablation Study}
	\subsubsection{The Effectiveness of Denoiser} 
\begin{figure}
		\centering
		\subfloat[]{ \includegraphics[width=0.22\textwidth]{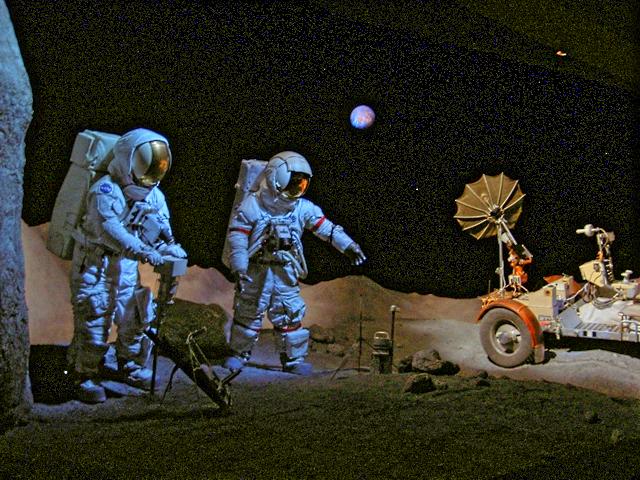}\label{a1}	
		} 
		\subfloat[]{ \includegraphics[width=0.22\textwidth]{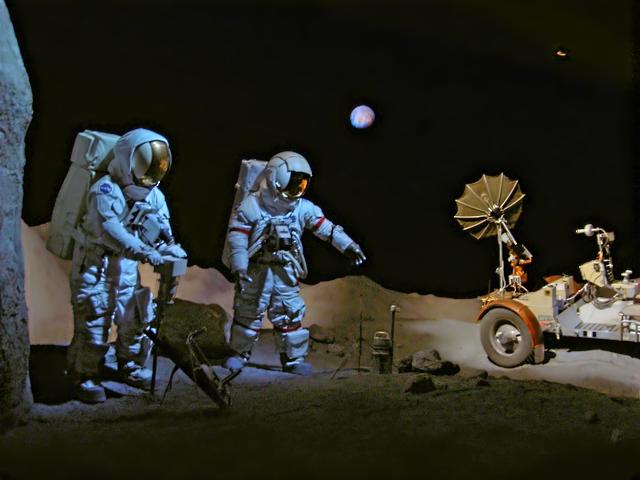}
		} 
		\caption{The results of our proposed framework with and without the denoiser. (a) W/O denoiser (NIQE: 3.6909), (b) with denoiser (NIQE: 2.8494).}
		\label{AS}
	\end{figure}

	To show the necessity of the inserted denoiser, we compare the results of our model with and without denoiser.
	Fig. \ref{AS} reports the comparable results.
	From it, we can find that the result generated by our proposed model without denoiser has heavy noise, color distortion and poor numerical result.
	In contrast, the results created by the model with denoiser have a clear and sharper structure.
	Such results have demonstrated the effectiveness of the proposed plug-and-play framework with the inserted denoisers, which can alleviate color bias and structural detail loss.
	By increasing the coefficient of denoising, we can increase the degree of noise suppression in the reflectance map.
	According to the degree of image degradation, different coefficients of denoising can be employed to achieve desirable performance.
	\begin{figure}
		\centering

		\subfloat[]{ \includegraphics[width=0.115\textwidth]{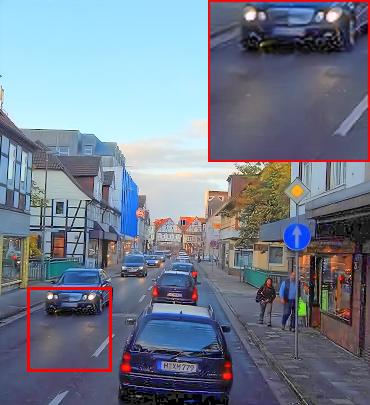}}
		\subfloat[]{ \includegraphics[width=0.115\textwidth]{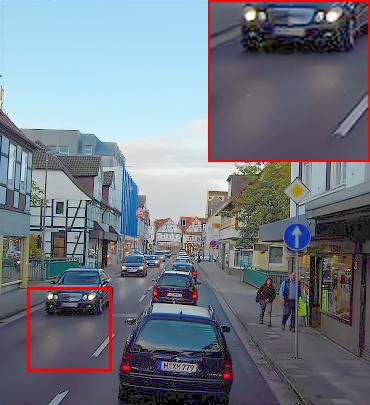}} 
		\subfloat[]{ \includegraphics[width=0.115\textwidth]{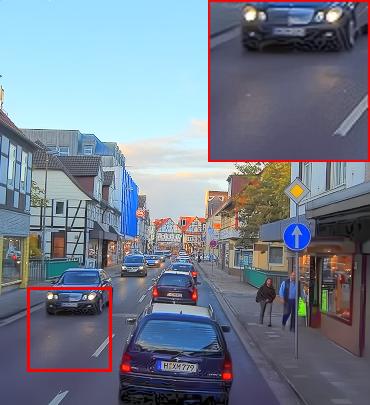}}  
		\subfloat[]{ \includegraphics[width=0.115\textwidth]{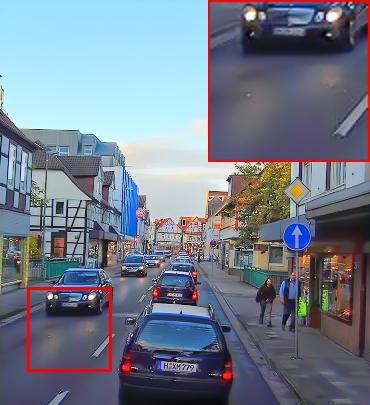}} 
		\caption{Performance of the proposed framework with different denoisers. (a) CBDNet (NIQE: 5.6149). (b) FFDNet (NIQE: 6.1877). (c) IRCNN (NIQE: 4.8271). (d) \ff{Our CNN-based denoiser (NIQE: 4.7822)}.}
		\label{denoisers} 
	\end{figure}	
 
	\subsubsection{Effects of Different CNN-based Denoisers} \label{denoisor}
	We compare the enhancement effects of different denoisers to prove the robustness of our plug-and-play framework. Fig. \ref{denoisers} shows the enhancement results of 4 different denoisers, including IRCNN \cite{zhang2017learning}, FFDNet \cite{zhang2018ffdnet}, CBDNet \cite{guo2019toward} and our CNN-based denoiser. From Fig. \ref{denoisers}, among all the denoisers, we can find that our denoiser achieves the best performance on both the visual effect and numerical result. Obviously, the performance of denoisers has a direct effect on the result of our proposed framework. However, the proposed method is a universal, feasible and superior framework, which allows the insertion of different denoisers based on the intrinsic characteristics of different images. 
	
 \subsubsection{Effect of An Explainable Denoiser} 
\begin{figure}
		\centering
		\subfloat[]{ \includegraphics[height=0.125\textwidth]{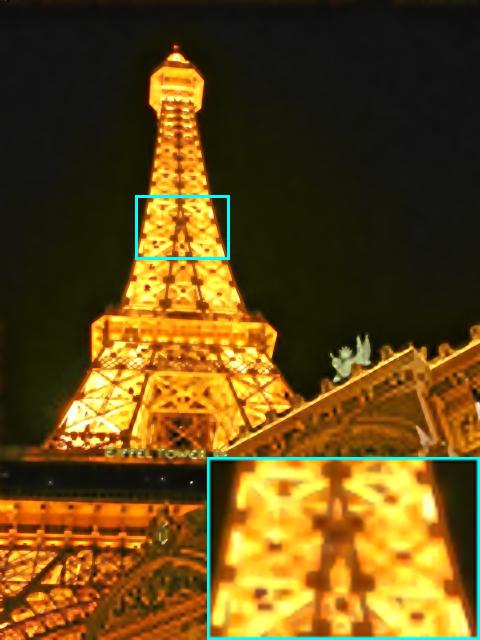}}  
		\subfloat[]{ \includegraphics[height=0.125\textwidth]{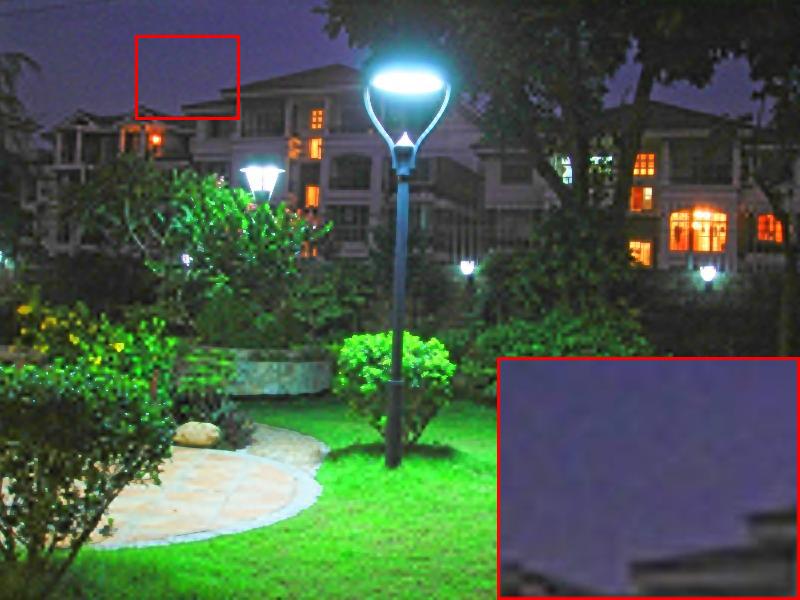}}  
		\subfloat[]{ \includegraphics[height=0.125\textwidth]{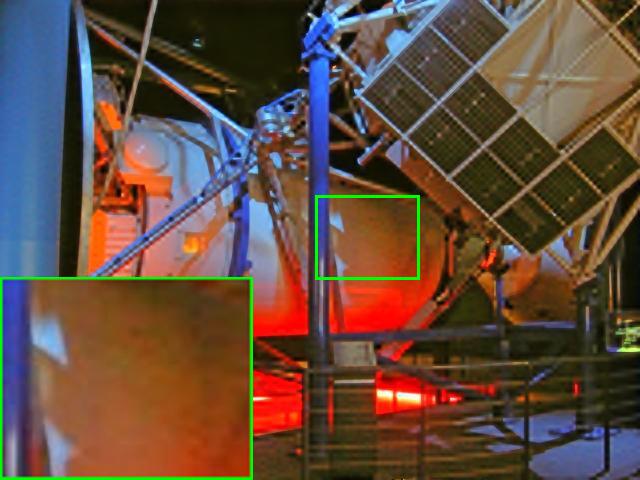}}  \\	
		\subfloat[]{ \includegraphics[height=0.125\textwidth]{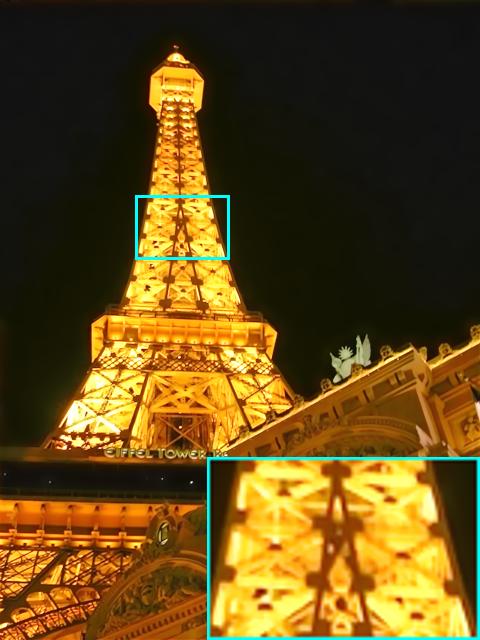}} 
		\subfloat[]{ \includegraphics[height=0.125\textwidth]{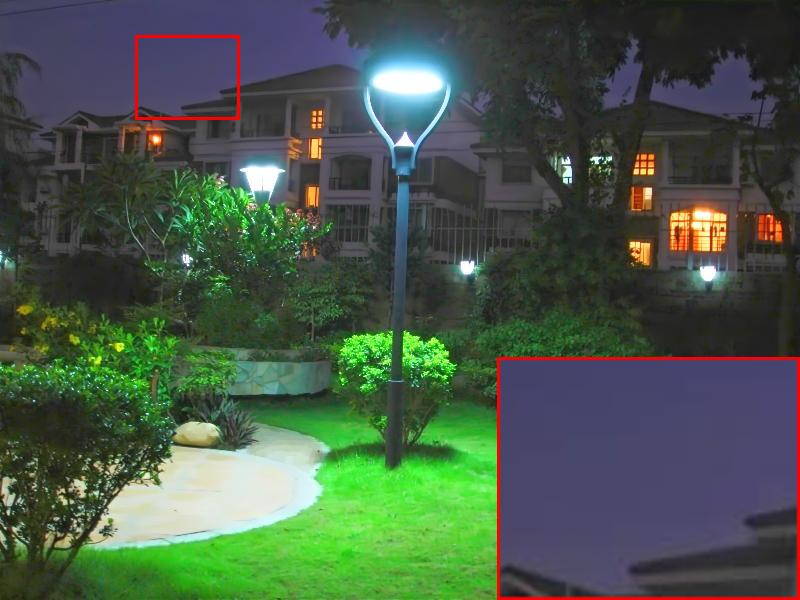}} 
		\subfloat[]{ \includegraphics[height=0.125\textwidth]{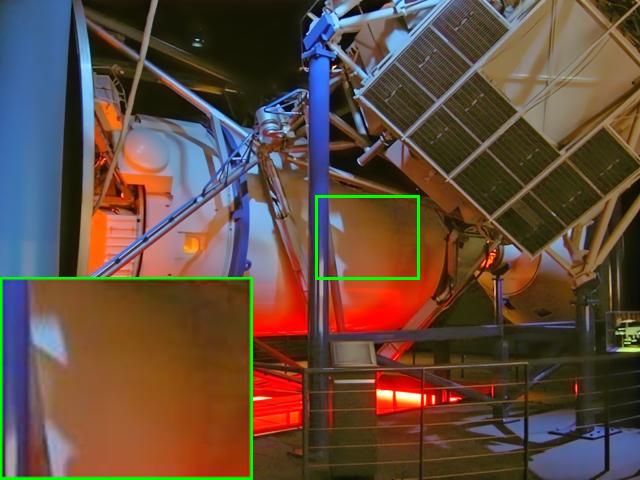}} 
		\caption{Visual comparison on wavelet denoiser (top row) and ours (bottom row).} 
		\label{wave}
	\end{figure}

\ff{Especially, we compare the effect of wavelet shrinkage denoiser (top row) and the Soft-AE (bottom row) with details in Fig. \ref{wave}. The denoised results of wavelet shrinkage denoiser still suffer obvious punctuate noise (the enlarged area in the red and green boxes) and lose structural details (the enlarged area in the blue box). On the contrary, the Soft-AE satisfactorily suppresses the noise and simultaneously keeps structural details. Thus, we conclude that the Soft-AE has better performance.}

 \subsection{Parametric Analysis}
 \ff{More favorably, for important hyperparameters such as $\alpha, \beta, \lambda, \gamma_1, \gamma_2$ which greatly affect regularization and correction, we have discussed their sensitivity to the performance of the proposed framework in detail.}
	\begin{figure*}[t!]
		\centering
		\subfloat[]{   \includegraphics[width=0.23\textwidth]{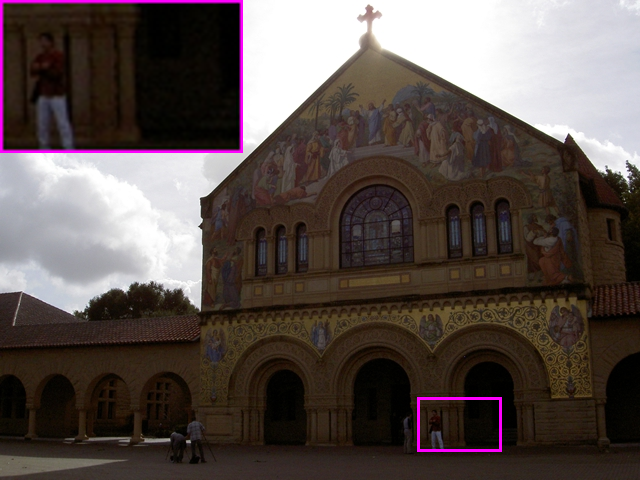}} 
		\subfloat[]{   \includegraphics[width=0.23\textwidth]{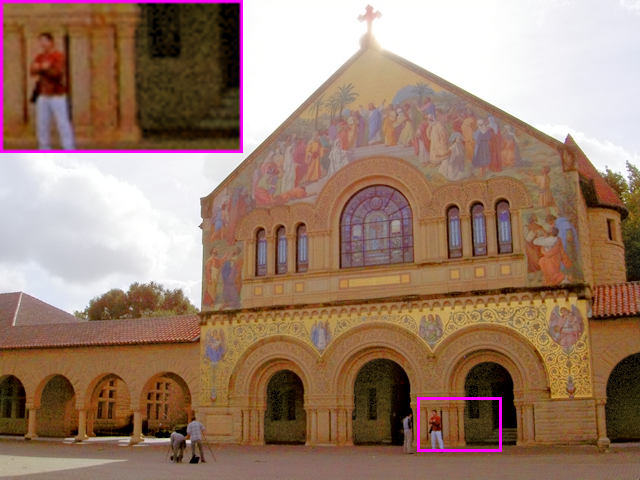}} 
		\subfloat[]{	\includegraphics[width=0.23\textwidth]{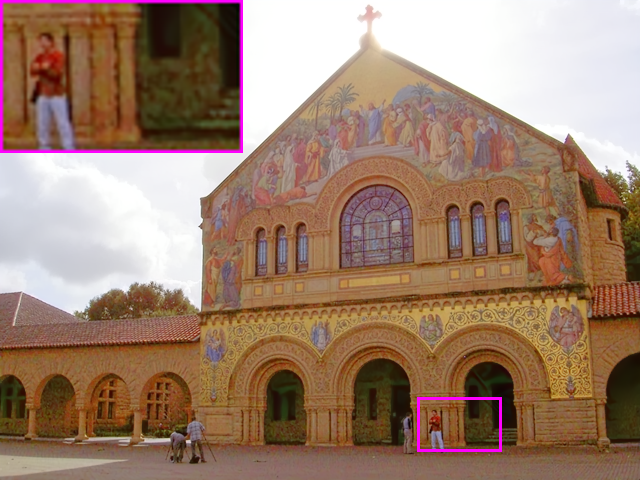}}
		\subfloat[]{	\includegraphics[width=0.23\textwidth]{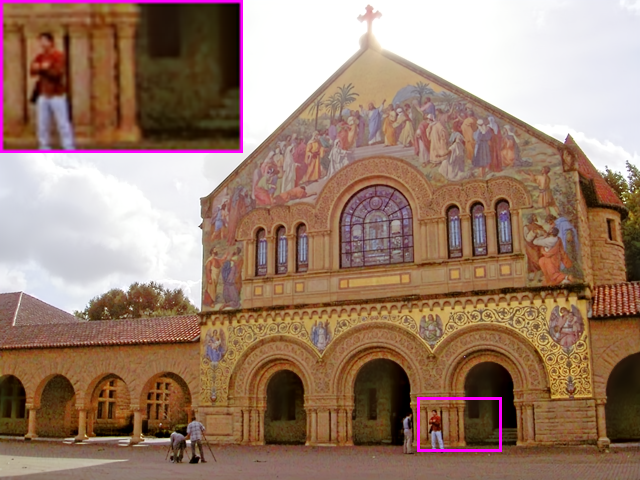}}
		\caption{Regularization parametric analysis. (a) input. (b) enhanced image with $\alpha=0$. (c) enhanced image with $\beta=0$. (d) enhanced image by our framework.}
		\label{p}
	\end{figure*}	
	\subsubsection{Regularization Parameters -- $\alpha$, $\beta$}
\begin{figure}
		\centering
		\subfloat[$\alpha$]{ \includegraphics[width=0.22\textwidth]{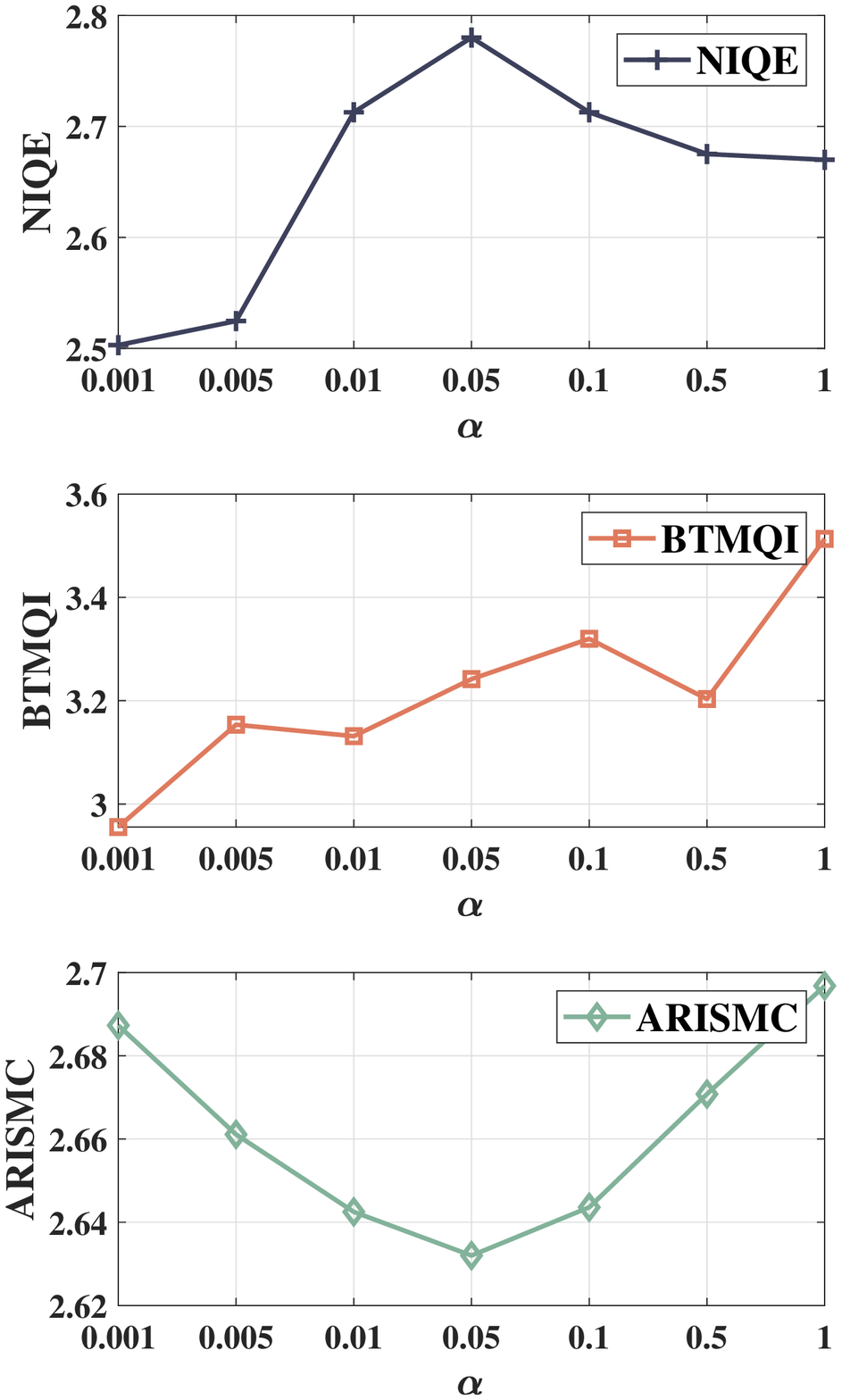}} 
		\subfloat[$\beta$]{ \includegraphics[width=0.22\textwidth]{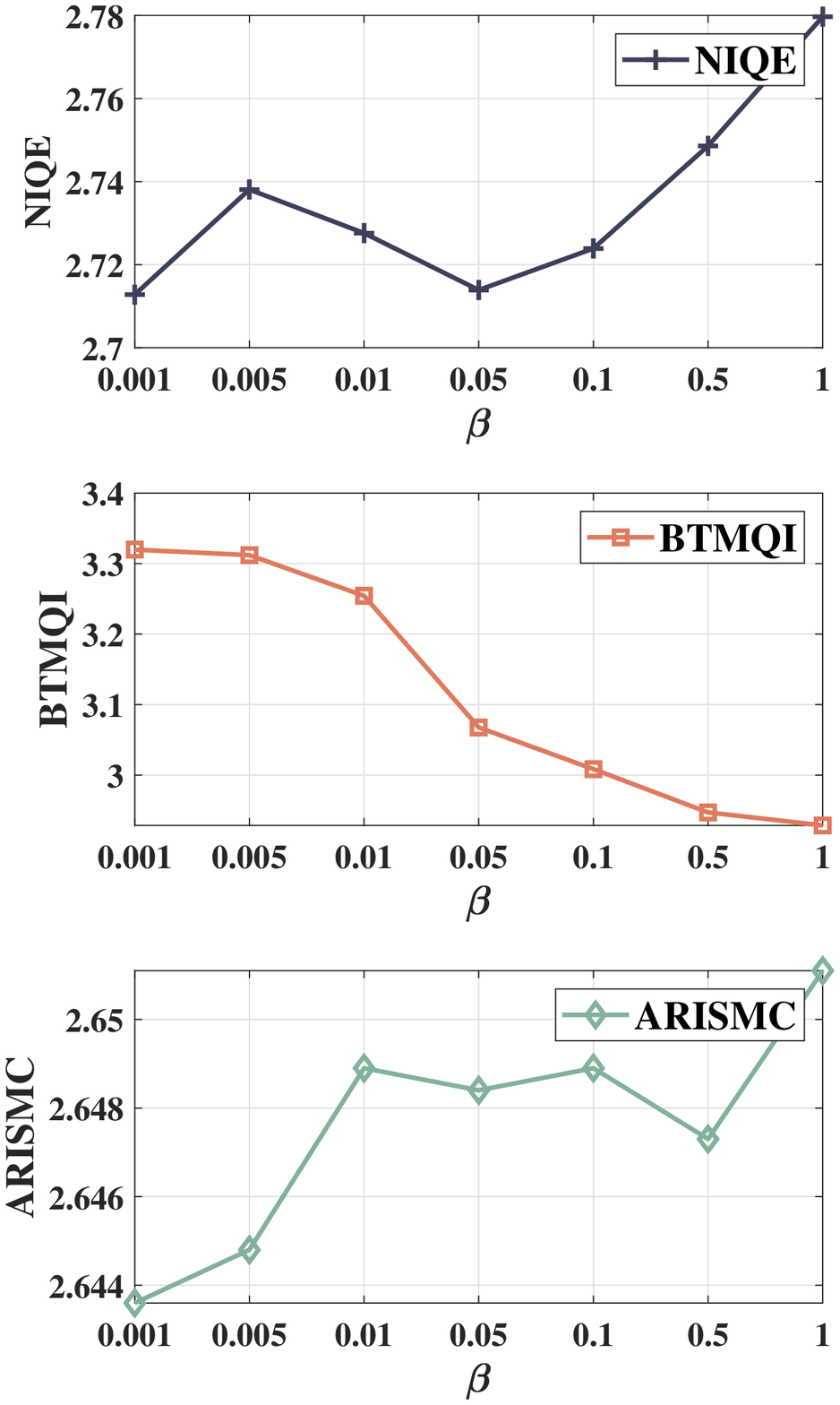}} 
		\caption{The effect of different regularization parameters on NIQE, BTQMI and ARISMC. We fix $\beta=0.001$ in (a), $\alpha=0.1$ in (b).}
		\label{par} 
	\end{figure}

	The regularization parameters $\alpha$ and $\beta$ in Eq. (\ref{R}) have direct effects on the performance of our model. Fig. \ref{p} shows the visual effects under different settings of $\alpha$ and $\beta$. Fig. \ref{p}(b) shows the result with $\alpha=0,$ which means the term $\Vert\nabla L\Vert_1$ does not work. Such a setting is not conducive to generating a smooth illumination map and leads to unsatisfactory enhancement results.
	Fig. \ref{p}(c) gives the result with $\beta=0$, which means the third term of Eq. (\ref{R}) is inactive.
	Fig. \ref{p}(d) is the result generated by our model with $\alpha=0.02$ and $\beta=0.001$. Such results demonstrate the usefulness of the second and third terms in Eq. (\ref{R}). To further test the effects of $\alpha$ and $\beta$, we conduct the experiments by only changing one parameter at a time. Fig. \ref{par} gives experimental results (NIQE, BTMQI and ARISMC) on Set12. From it, we can easily observe that different parameters have a direct impact on the performance of our method. Please note that lower NIQE, BTMQI and ARISMC values represent better visual quality. As can be observed, NIQE always prefers smaller parameters but is not sensitive to $\beta$. BTMQI prefers smaller $\alpha$ and larger $\beta$. ARISMC prefers an intermediate $\alpha$ and is not sensitive to $\beta$.
	\begin{figure}
		\centering
		
		\subfloat[Noise level 15]{ \includegraphics[height=0.14\textwidth]{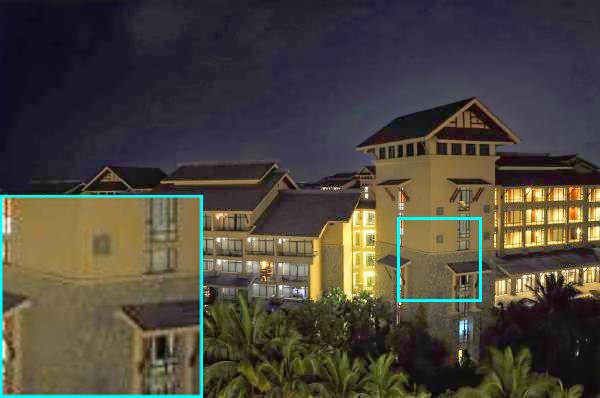}} 
		\subfloat[Noise level 25]{ \includegraphics[height=0.14\textwidth]{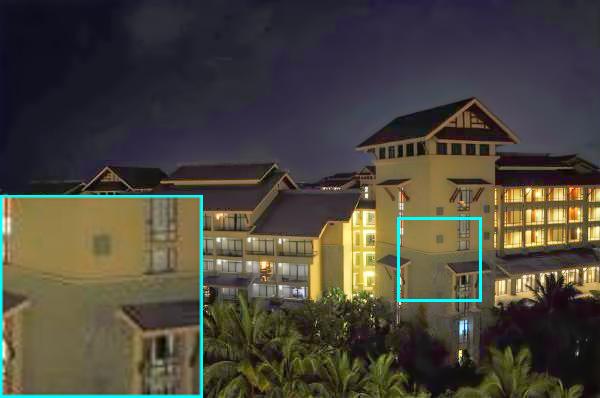}} \\
		\subfloat[]{ \includegraphics[height=0.18\textwidth]{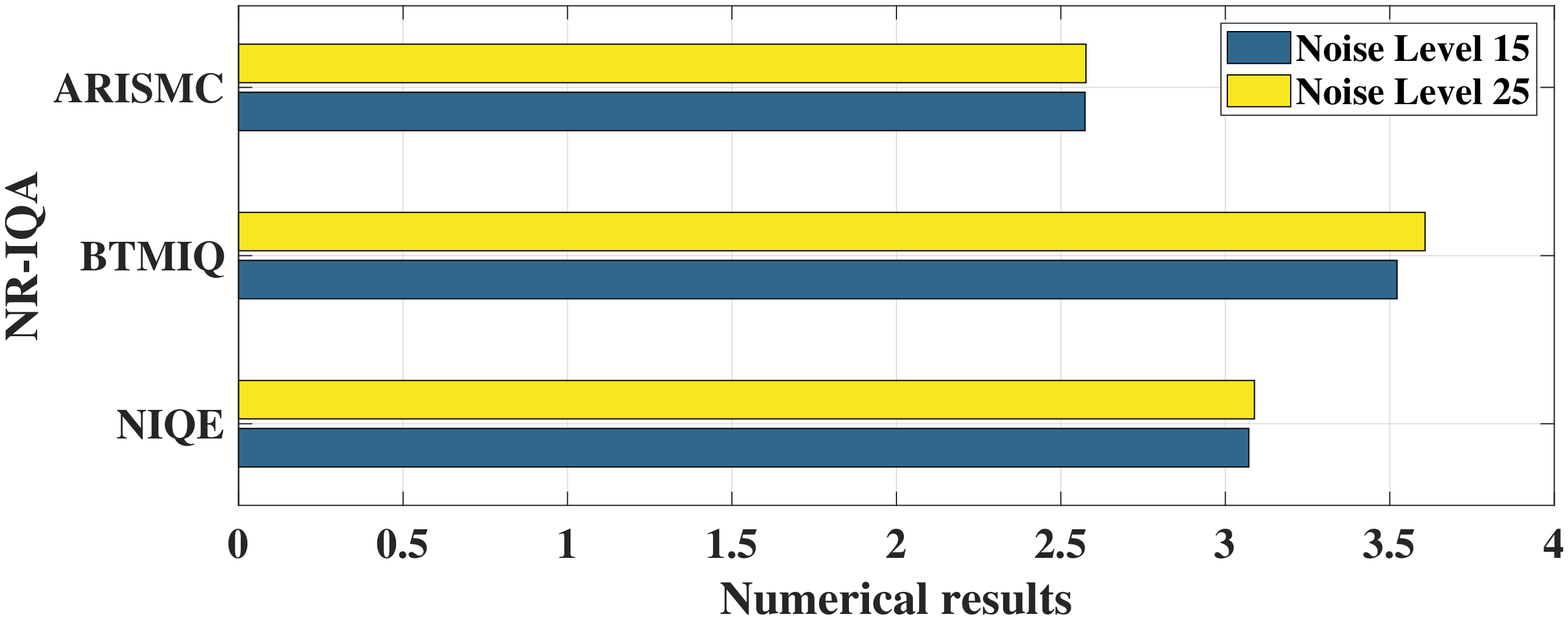}} \\
		
		\caption{\ff{The quantitative and qualitative analysis of different $\lambda$.}}
		\label{lambda} 
	\end{figure}
	\subsubsection{\ff{The noise level -- $\sqrt{\lambda}$}}
	\ff{The noise level of the denoiser usually determines the effect of denoising. Therefore, we studied the effect of different denoising levels on the enhanced results. Fig. \ref{lambda} demonstrates the quantitative and qualitative analysis of different $\lambda$. It is obvious that the CNN-based denoiser with noise level 25 achieves a better visual result. However, the enhanced results obtained by the denoisers with different noise levels have little difference in numerical results. In order to pursue a better denoising performance, we tend to use a CNN-based denoiser with noise level 25.}
	\subsubsection{Gamma Correction: $\gamma_1$, $\gamma_2$}
\begin{figure}[t]
		\centering
		\subfloat[PSNR]{ \includegraphics[width=0.22\textwidth]{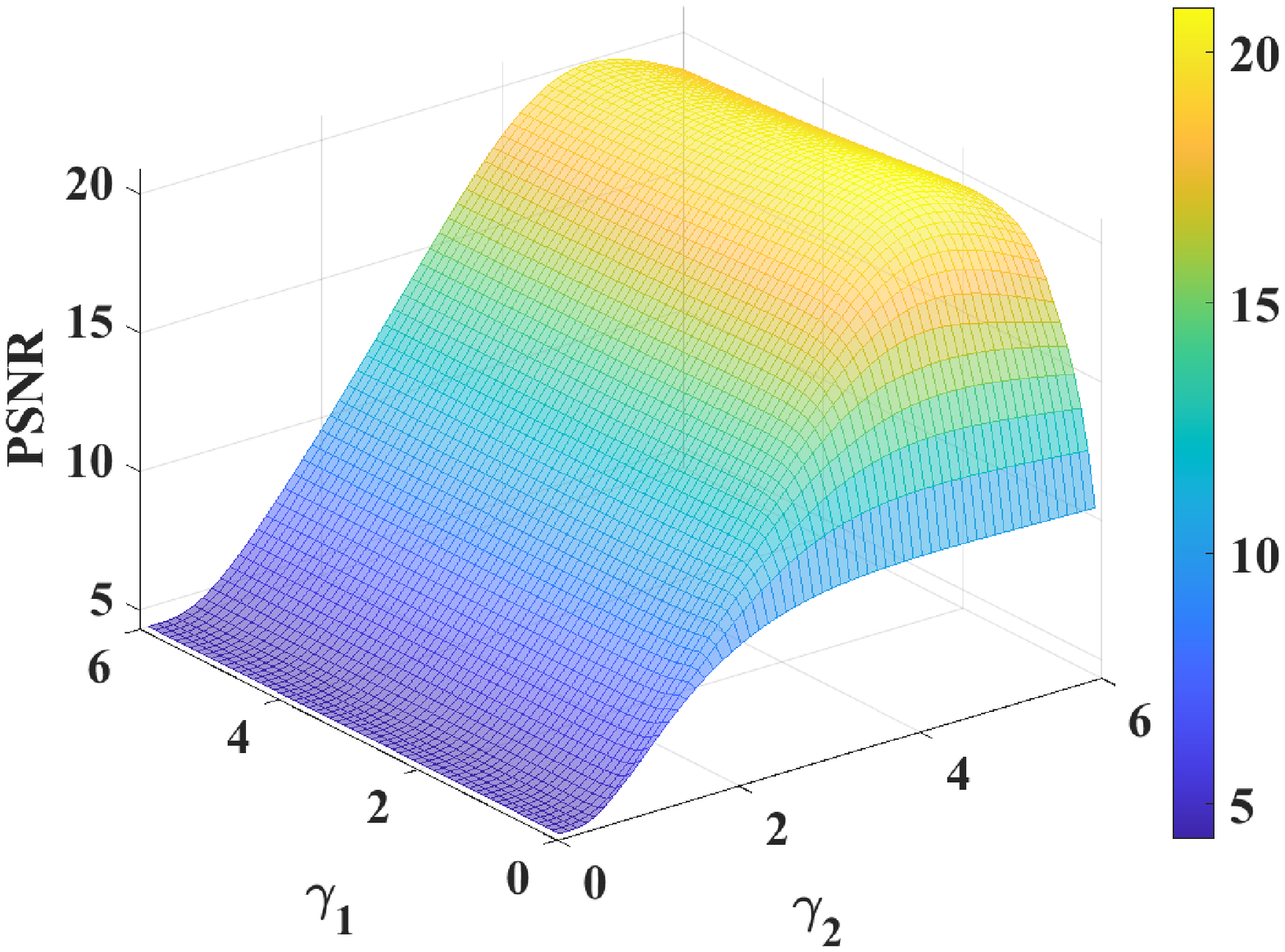}} 
		\subfloat[SSIM]{ \includegraphics[width=0.22\textwidth]{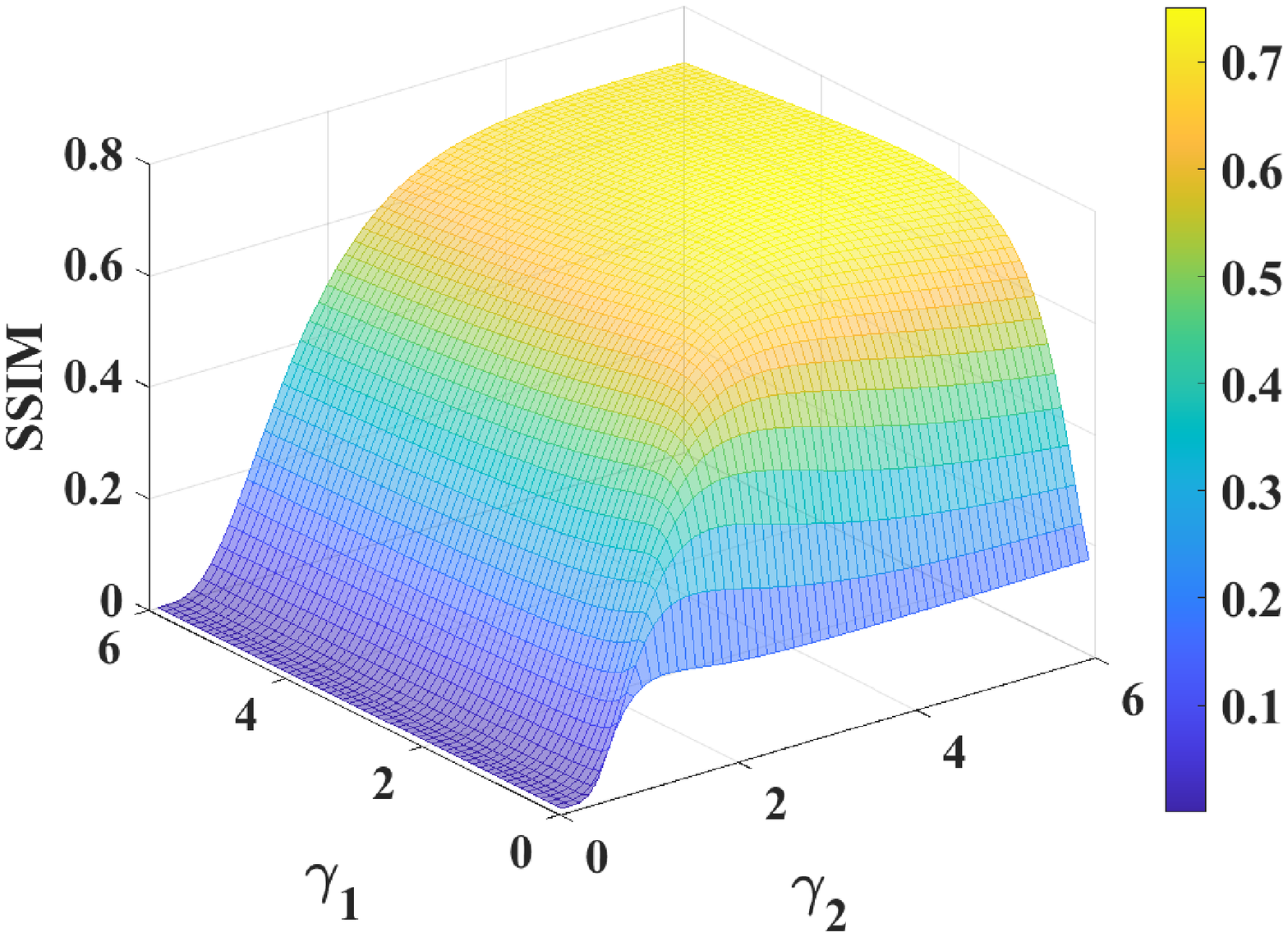}}\\
		\subfloat[MSE]{ \includegraphics[width=0.23\textwidth]{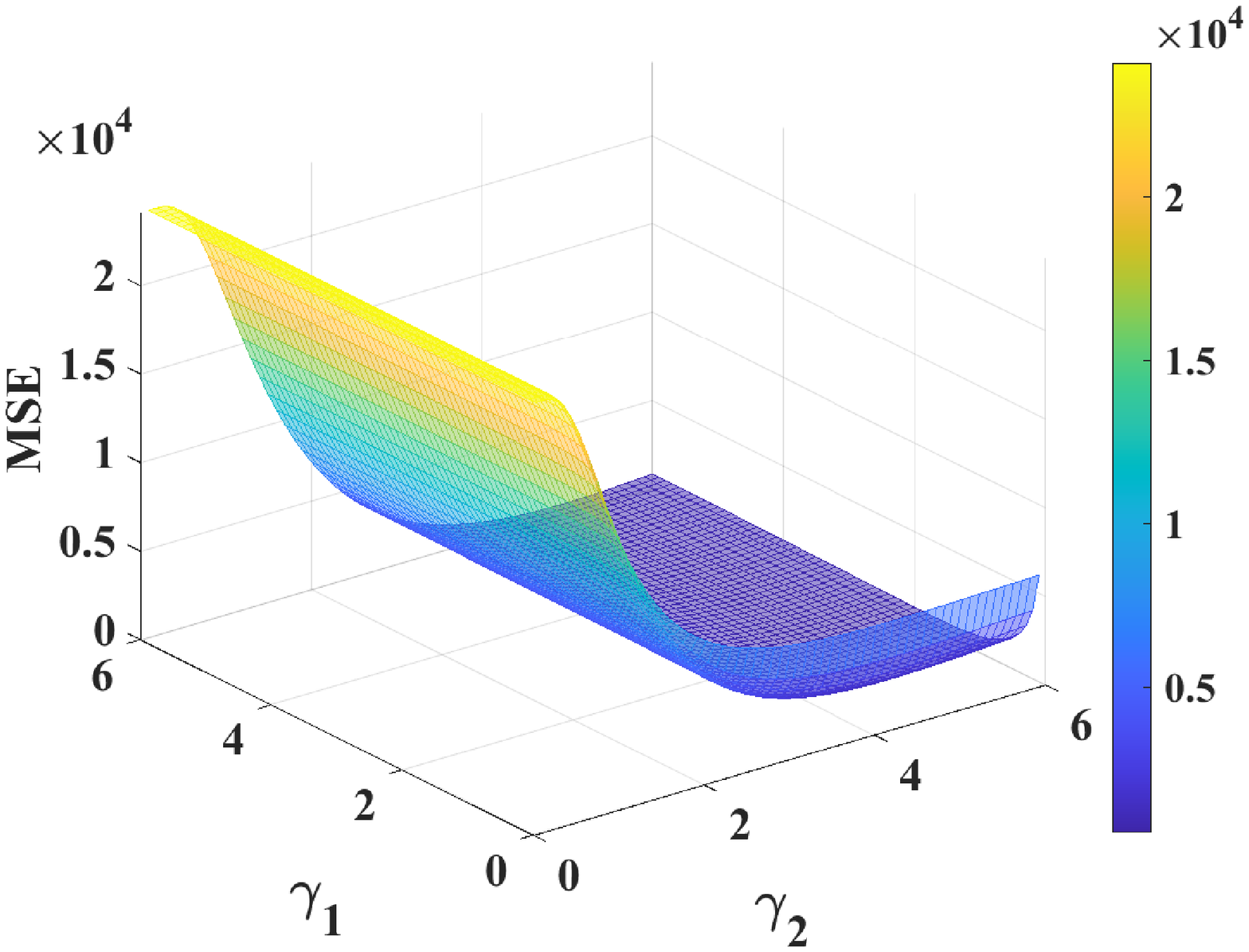}} 
		\subfloat[NIQE]{ \includegraphics[width=0.22\textwidth]{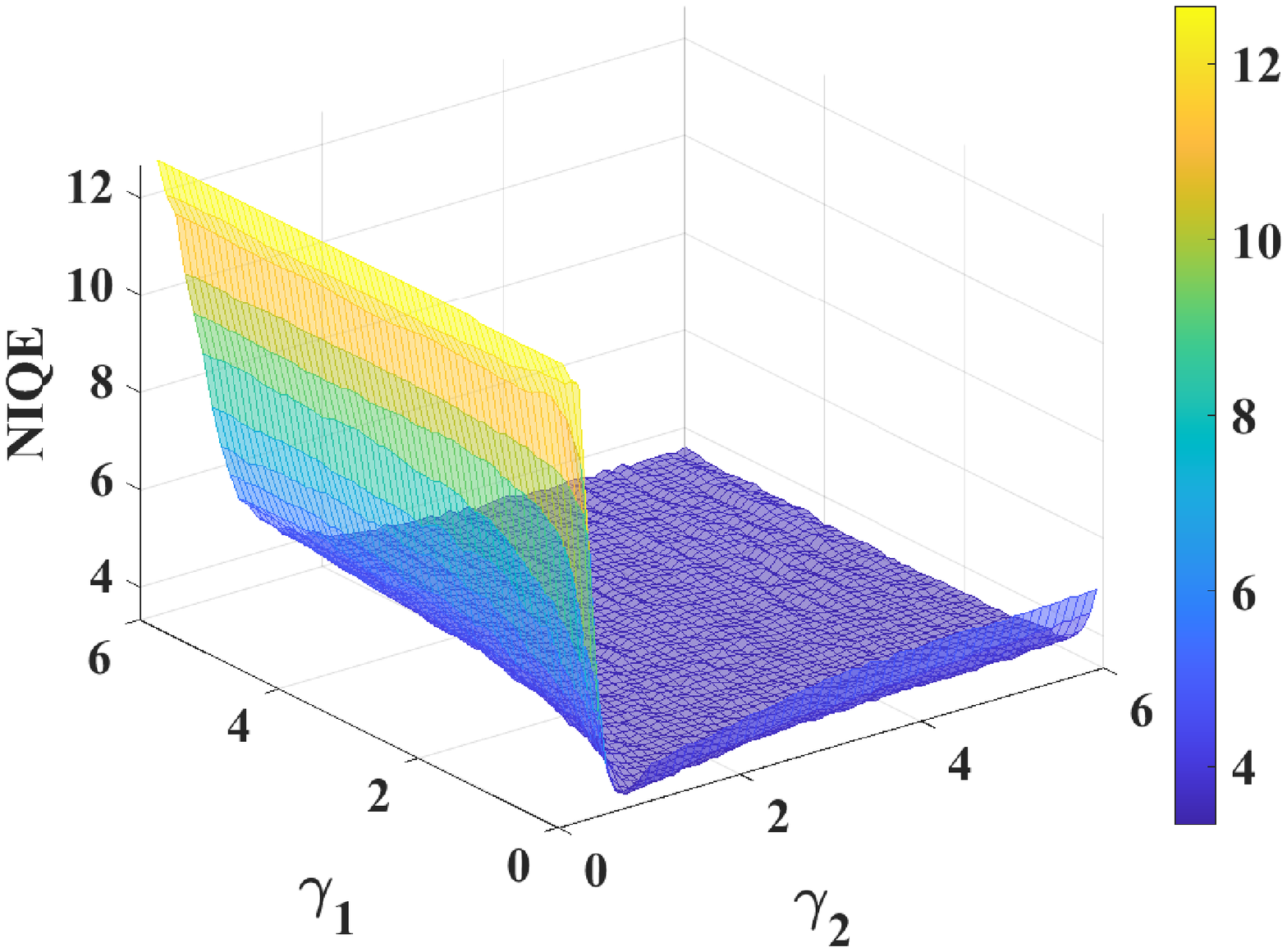}} 
		\caption{Impact of gamma correction parameters on PSNR, SSIM, MSE, and NIQE.}
		\label{gammapa} 
	\end{figure}

	Gamma correction is a non-negligible link in the Retinex method. 
	Fig. \ref{gammapa} demonstrates the numerical results with different $\gamma_1$ and $\gamma_2$ on the test set of LOL. The result indicates that the numerical results are highly affected by $\gamma_2$. A lower $\gamma_2$ value corresponds to a higher NIQE value and a higher MSE value. Moreover, PSNR and SSIM are more inclined to a higher $\gamma_2$. On the contrary, the performance is not sensitive to the variation of $\gamma_1$. Therefore, to obtain stable and superior performance in low-light image enhancement, we adopt different settings of $\gamma_1$ and $\gamma_2$ on different datasets. For the LOL dataset, $\gamma_1$ and $\gamma_2$ are set to 1.5 and 4, respectively, while $\gamma_1=1$ and $\gamma_2=2.2$ are set on Set12 dataset.
	
	\section{Conclusion} \label{conclusion}
	\ff{In this paper, we have proposed a framework that uses a sequential decomposition strategy to solve $R$ and $L$ sequentially, thereby avoiding the alternating iteration and canceling the mutual interference between solving $R$ and $L$. This framework not only circumvents the reliance on large paired low/normal-light data, a key problem encountered in image enhancement, but also promotes interpretability. Quantitative and qualitative experiments have demonstrated the superiority of our method. However, we find that the visual effect of images is largely affected by the gamma correction parameters which depend on the degree of underexposure. Future work will be investigating the effect of gamma correction and exploring efficient gamma correction methods.}
	
	
	
	\bibliographystyle{IEEEtran}
	\bibliography{ref}

	\newpage
	\vfill
\end{document}